%% file: paper.tex
\documentclass[sigconf]{acmart}
\usepackage{booktabs} % For formal tables
\usepackage{todonotes}
\usepackage{microtype}
\usepackage{xspace}
\usepackage{balance}
\usepackage[english]{babel}
\usepackage[capitalise]{cleveref}
\usepackage{subfig}
\usepackage{csquotes}
\usepackage{multicol}
\usepackage{multirow}
\usepackage{xstring}
\usepackage{dcolumn}

\DeclareQuoteStyle[american]{english}
        {\itshape\textquotedblleft}
        [\textquotedblleft]
        {\textquotedblright}
        [0.05em]
        {\textquoteleft}
        {\textquoteright}

\newcolumntype{d}{D{.}{.}{2.3}}
\newcolumntype{e}{D{.}{.}{4.0}}

\newcommand{\OR}{$OR $}
\AtBeginDocument{%
  \providecommand\BibTeX{{%
    \normalfont B\kern-0.5em{\scshape i\kern-0.25em b}\kern-0.8em\TeX}}}

\copyrightyear{2021}
\acmYear{2021}
\setcopyright{acmcopyright}\acmConference[WiPSCE '21]{The 16th Workshop in Primary and Secondary Computing Education}{October 18--20, 2021}{Virtual Event, Germany}
\acmBooktitle{The 16th Workshop in Primary and Secondary Computing Education (WiPSCE '21), October 18--20, 2021, Virtual Event, Germany}
\acmPrice{15.00}
\acmDOI{10.1145/3481312.3481344}
\acmISBN{978-1-4503-8571-8/21/10}

\newcommand{\summary}[2]{
        \vspace{0.4em}
        \noindent
        \colorbox{gray!20}{%
            \parbox{.97\linewidth}{%
                    \textbf{\textsf{Summary (\textit{#1})}}
                #2
            }%
        }%
}%

\begin{document}

\input{macros}
\input{stats.tex}

\title[Effects of Hints on Debugging Scratch Programs]{Effects of Hints on Debugging Scratch Programs: \\An Empirical Study with Primary School Teachers in Training}
%\title{Effects of Hints on Debugging Scratch Programs}
%\title{Hints on Bug Patterns to Support Debugging Scratch Programs}


\author{Luisa Greifenstein}
\email{luisa.greifenstein@uni-passau.de}
\affiliation{%
	\institution{University of Passau}
	\city{Passau}%
	\country{Germany}
}

\author{\mbox{Florian Obermüller}}
\email{florian.obermueller@uni-passau.de}
\affiliation{%
	\institution{University of Passau}
	\city{Passau}
	\country{Germany}
}

\author{Ewald Wasmeier}
\email{ewald.wasmeier@uni-passau.de}
\affiliation{%
	\institution{University of Passau}
	\city{Passau}%
	\country{Germany}
}

\author{Ute Heuer}
\email{ute.heuer@uni-passau.de}
\affiliation{%
	\institution{University of Passau}
	\city{Passau}%
	\country{Germany}
}

\author{Gordon Fraser}
\email{gordon.fraser@uni-passau.de}
\affiliation{%
	\institution{University of Passau}
	\city{Passau}
	\country{Germany}
}

\renewcommand{\shortauthors}{Greifenstein et al.}

\begin{abstract}
Bugs in learners' programs are often the result of fundamental misconceptions.
Teachers frequently face the challenge of first having to understand such bugs, and then suggest ways to fix them. 
In order to enable teachers to do so effectively and efficiently, it is desirable to support them in recognising and fixing bugs.
Misconceptions often lead to recurring patterns of similar bugs, enabling automated tools to provide this support in terms of hints on occurrences of common bug patterns.
In this paper, %we study the effects of providing teachers such hints for bugs in learners' programs. 
we investigate to what extent the hints improve the effectiveness and efficiency of teachers in debugging learners' programs using a cohort of \numberOfParticipants~primary school teachers in training, tasked to correct buggy \scratch programs, with and without hints on bug patterns. Our experiment suggests that automatically generated hints can reduce the effort of finding and fixing bugs from \meanProcessingTimeCtrlLower~to \meanProcessingTimeTrmtLower~minutes%by \meanProcessingTimeDifferenceLower~minutes
, while increasing the effectiveness by  \percentageFunctionalityDifferenceLower\% more correct solutions.
While this improvement is convincing, arguably teachers in training might first need to learn debugging ``the hard way'' to not miss the opportunity to learn by relying on tools. We therefore investigate whether the use of hints during training affects their ability to recognise and fix bugs without hints. 
Our experiment provides no significant evidence that either learning to debug with hints or learning to debug ``the hard way'' leads to better learning effects.
% although we observe a slight improvement of having learnt with hints. %in terms of efficiency based on the recognition of recurring bug patterns
%
Overall, this suggests that bug patterns might be a useful concept to include in the curriculum for teachers in training, while tool-support to recognise these patterns is desirable for teachers in practice.
\end{abstract}




\newcommand{\litterbox}{\textsc{LitterBox}\xspace}
\newcommand{\scratch}{\textsc{Scratch}\xspace}
\newcommand{\drscratch}{\textsc{Dr. Scratch}\xspace}
\newcommand{\hairball}{\textsc{Hairball}\xspace}
\newcommand{\qualityhound}{\textsc{Quality Hound}\xspace}
\newcommand{\findbugs}{\textsc{FindBugs}\xspace}
\newcommand{\whisker}{\textsc{Whisker}\xspace}

\maketitle

\input{content/introduction}
\input{content/relatedwork}

\input{content/study}
\input{content/results}
\input{content/discussion}

\input{content/conclusions}

\vspace{-0.3em}
\begin{acks}
\vspace{-0.3em}This work is supported by the Federal Ministry of Education and Research
through project ``primary::programming'' (01JA2021) as
part of the ``Qualitätsoffensive Lehrerbildung'', a joint initiative of the
Federal Government and the Länder. The authors are responsible for the content
of this publication.
\end{acks}

\bibliographystyle{ACM-Reference-Format}
\bibliography{references}

\end{document}

%% file: macros.tex
\newcommand{\numLectureParticipants}
{242\xspace}

\newcommand{\droupoutsTotal}{26}
\newcommand{\dropoutsCtrl}{16\xspace}
\newcommand{\dropoutsTrmt}{10\xspace}
\newcommand{\dropoutsCtrlNotAllTasks}{11\xspace}
\newcommand{\dropoutsTrmtNotAllTasks}{7\xspace}
\newcommand{\dropoutsCtrlFalseSubmissions}{5\xspace}
\newcommand{\dropoutsTrmtFalseSubmissions}{3\xspace}

\newcommand{\groupA}{Group \textit{Ctrl}\xspace}
\newcommand{\groupB}{Group \textit{Trmt}\xspace}

\newcommand{\numDeviationsRoundOne}{13}

%%% Hint Evaluation
\newcommand{\hintEvalPositiveTaskI}{87\xspace}
\newcommand{\hintEvalPositiveTaskII}{115\xspace}
\newcommand{\hintEvalPositiveTaskIII}{91\xspace}
\newcommand{\hintEvalPositiveTaskIV}{90\xspace}
\newcommand{\hintEvalPositiveTaskV}{70\xspace}
\newcommand{\hintEvalPositiveTaskVI}{86\xspace}
\newcommand{\hintEvalPositiveTaskVII}{77\xspace}

\newcommand{\hintEvalNeutralTaskI}{3\xspace}
\newcommand{\hintEvalNeutralTaskII}{5\xspace}
\newcommand{\hintEvalNeutralTaskIII}{3\xspace}
\newcommand{\hintEvalNeutralTaskIV}{7\xspace}
\newcommand{\hintEvalNeutraleTaskV}{11\xspace}
\newcommand{\hintEvalNeutralTaskVI}{3\xspace}
\newcommand{\hintEvalNeutralTaskVII}{6\xspace}

\newcommand{\hintEvalNegativeTaskI}{13\xspace}
\newcommand{\hintEvalNegativeTaskII}{8\xspace}
\newcommand{\hintEvalNegativeTaskIII}{18\xspace}
\newcommand{\hintEvalNegativeTaskIV}{22\xspace}
\newcommand{\hintEvalNegativeTaskV}{41\xspace}
\newcommand{\hintEvalNegativeTaskVI}{8\xspace}
\newcommand{\hintEvalNegativeTaskVII}{28\xspace}

\newcommand{\hintEvalUnnecessaryTaskI}{6\xspace}
\newcommand{\hintEvalUnnecessaryTaskII}{4\xspace}
\newcommand{\hintEvalUnnecessaryTaskIII}{8\xspace}
\newcommand{\hintEvalUnnecessaryTaskIV}{3\xspace}
\newcommand{\hintEvalUnnecessaryTaskV}{8\xspace}
\newcommand{\hintEvalUnnecessaryTaskVI}{4\xspace}
\newcommand{\hintEvalUnnecessaryTaskVII}{2\xspace}

\newcommand\add[1]{\textcolor{red}{[[ #1]]}}

%% file: stats.tex
\newcommand{\interRaterReliabilityHintEvaluationKappa}{.75}
\newcommand{\numberOfParticipants}{163}
\newcommand{\numberOfActiveParticipants}{189}
\newcommand{\numberOfParticipantsCtrl}{78}
\newcommand{\numberOfParticipantsTrmt}{85}
\newcommand{\numberOfFemaleParticipants}{142}
\newcommand{\numberOfMaleParticipants}{21}
\newcommand{\ageYoungestParticipant}{18}
\newcommand{\ageOldestParticipant}{45}
\newcommand{\ageParticipantsMedian}{20}
\newcommand{\ageParticipantsMean}{20.47}
\newcommand{\numberOfExeriencedParticipants}{104}
\newcommand{\numberOfUnexeriencedParticipants}{59}
\newcommand{\numberOfExeriencedParticipantsCtrl}{51}
\newcommand{\numberOfUnexeriencedParticipantsCtrl}{27}
\newcommand{\numberOfExeriencedParticipantsTrmt}{53}
\newcommand{\numberOfUnexeriencedParticipantsTrmt}{32}
\newcommand{\percentageExeriencedParticipantsCtrl}{65}
\newcommand{\percentageExeriencedParticipantsTrmt}{62}
\newcommand{\percentageUnexeriencedParticipantsCtrl}{35}
\newcommand{\percentageUnexeriencedParticipantsTrmt}{38}
\newcommand{\percentageFunctionalityCtrlLower}{48}
\newcommand{\percentageFunctionalityCtrlHigher}{47}
\newcommand{\percentageFunctionalityDifferenceLower}{34}
\newcommand{\percentageFunctionalityTrmtLower}{82}
\newcommand{\percentageFunctionalityTrmtHigher}{54}
\newcommand{\percentageFunctionalityDifferenceHigher}{7}
\newcommand{\meanAmountBugPatternCtrlLower}{1.05}
\newcommand{\meanAmountBugPatternCtrlHigher}{1.18}
\newcommand{\meanAmountBugPatternTrmtLower}{.33}
\newcommand{\meanAmountBugPatternTrmtHigher}{.93}
\newcommand{\meanProcessingTimeCtrlLower}{8.66}
\newcommand{\meanProcessingTimeCtrlHigher}{7.69}
\newcommand{\meanProcessingTimeDifferenceLower}{3.41}
\newcommand{\meanProcessingTimeTrmtLower}{5.24}
\newcommand{\meanProcessingTimeTrmtHigher}{7.00}
\newcommand{\meanProcessingTimeDifferenceHigher}{.69}
\newcommand{\meanProcessingTimeJustCorrectCtrlLower}{6.81}
\newcommand{\meanProcessingTimeJustCorrectCtrlHigher}{6.15}
\newcommand{\meanProcessingTimeJustCorrectTrmtLower}{4.96}
\newcommand{\meanProcessingTimeJustCorrectTrmtHigher}{5.23}
\newcommand{\meanProcessingTimeJustWrongCtrlLower}{10.36}
\newcommand{\meanProcessingTimeJustWrongCtrlHigher}{9.07}
\newcommand{\meanProcessingTimeJustWrongTrmtLower}{6.53}
\newcommand{\meanProcessingTimeJustWrongTrmtHigher}{9.07}
\newcommand{\percentageProcessingTimeCtrlTrmtImprovementLower}{39}
\newcommand{\percentageFunctionalityCtrlTrmtImprovementLower}{71}
\newcommand{\percentageProcessingTimeCtrlTrmtImprovementHigher}{9}
\newcommand{\percentageFunctionalityCtrlTrmtImprovementHigher}{14}
\newcommand{\meanProcessingTimeExperienceCtrlLower}{8.69}
\newcommand{\meanProcessingTimeExperienceCtrlHigher}{7.65}
\newcommand{\meanProcessingTimeExperienceTrmtLower}{5.17}
\newcommand{\meanProcessingTimeExperienceTrmtHigher}{6.96}
\newcommand{\meanProcessingTimeNoExperienceCtrlLower}{8.58}
\newcommand{\meanProcessingTimeNoExperienceCtrlHigher}{7.77}
\newcommand{\meanProcessingTimeNoExperienceTrmtLower}{5.37}
\newcommand{\meanProcessingTimeNoExperienceTrmtHigher}{7.08}
\newcommand{\whiskerTestsNumberOfAssertsMean}{9}
\newcommand{\whiskerTestsNumberOfAssertsMedian}{7}
\newcommand{\whiskerTestsNumberOfAssertsMax}{17}
\newcommand{\whiskerTestsNumberOfAssertsMin}{4}
\newcommand{\meanHintEvaluationAllHints}{4.21}
\newcommand{\meanHintEvaluationTaskI}{4.19}
\newcommand{\meanHintEvaluationTaskII}{4.51}
\newcommand{\meanHintEvaluationTaskIII}{4.20}
\newcommand{\meanHintEvaluationTaskIV}{4.12}
\newcommand{\meanHintEvaluationTaskV}{3.80}
\newcommand{\meanHintEvaluationTaskVI}{4.48}
\newcommand{\meanHintEvaluationTaskVII}{4.18}
\newcommand{\meanBlockCount}{53.00}
\newcommand{\meanCtScoreAverage}{1.83}
\newcommand{\meanHalsteadDifficulty}{18.76}
\newcommand{\meanHalsteadEffort}{9372.50}
\newcommand{\meanHalsteadLength}{85.14}
\newcommand{\meanHalsteadSize}{48.64}
\newcommand{\meanHalsteadVolume}{480.14}
\newcommand{\meanInterproceduralCyclomaticComplexity}{15.14}
\newcommand{\percentageCorrectSolutionsCtrlTaskI}{68}
\newcommand{\percentageCorrectSolutionsTrmtTaskI}{82}
\newcommand{\percentageCorrectSolutionsCtrlTaskII}{58}
\newcommand{\percentageCorrectSolutionsTrmtTaskII}{94}
\newcommand{\percentageCorrectSolutionsCtrlTaskIII}{49}
\newcommand{\percentageCorrectSolutionsTrmtTaskIII}{68}
\newcommand{\percentageCorrectSolutionsCtrlTaskIV}{10}
\newcommand{\percentageCorrectSolutionsTrmtTaskIV}{84}
\newcommand{\percentageCorrectSolutionsCtrlTaskV}{88}
\newcommand{\percentageCorrectSolutionsTrmtTaskV}{79}
\newcommand{\percentageCorrectSolutionsCtrlTaskVI}{47}
\newcommand{\percentageCorrectSolutionsTrmtTaskVI}{88}
\newcommand{\percentageCorrectSolutionsCtrlTaskVII}{15}
\newcommand{\percentageCorrectSolutionsTrmtTaskVII}{80}
\newcommand{\percentageCorrectSolutionsCtrlTaskVIII}{68}
\newcommand{\percentageCorrectSolutionsTrmtTaskVIII}{76}
\newcommand{\percentageCorrectSolutionsCtrlTaskIX}{67}
\newcommand{\percentageCorrectSolutionsTrmtTaskIX}{69}
\newcommand{\percentageCorrectSolutionsCtrlTaskX}{40}
\newcommand{\percentageCorrectSolutionsTrmtTaskX}{52}
\newcommand{\percentageCorrectSolutionsCtrlTaskXI}{18}
\newcommand{\percentageCorrectSolutionsTrmtTaskXI}{26}
\newcommand{\percentageCorrectSolutionsCtrlTaskXII}{36}
\newcommand{\percentageCorrectSolutionsTrmtTaskXII}{33}
\newcommand{\percentageCorrectSolutionsCtrlTaskXIII}{77}
\newcommand{\percentageCorrectSolutionsTrmtTaskXIII}{78}
\newcommand{\percentageCorrectSolutionsCtrlTaskXIV}{26}
\newcommand{\percentageCorrectSolutionsTrmtTaskXIV}{42}
\newcommand{\tasksLowerEffectSize}{5.00}
\newcommand{\tasksLowerChiSquarePValue}{p<.001}
\newcommand{\tasksHigherEffectSize}{1.30}
\newcommand{\tasksHigherChiSquarePValue}{p=.054}
\newcommand{\tasksLowerExperienceEffectSize}{5.36}
\newcommand{\tasksLowerExperienceChiSquarePValue}{p<.001}
\newcommand{\tasksHigherExperienceEffectSize}{1.19}
\newcommand{\tasksHigherExperienceChiSquarePValue}{p=.35}
\newcommand{\tasksLowerNoExperienceEffectSize}{4.76}
\newcommand{\tasksLowerNoExperienceChiSquarePValue}{p<.001}
\newcommand{\tasksHigherNoExperienceEffectSize}{1.61}
\newcommand{\tasksHigherNoExperienceChiSquarePValue}{p=.031}
\newcommand{\percentageFiveOrMoreCorrectTasksCtrlLower}{22}
\newcommand{\percentageFiveOrMoreCorrectTasksTrmtLower}{85}
\newcommand{\numberOfCorrectSolutionsGreaterThanFourLowerEffectSize}{19.87}
\newcommand{\numberOfCorrectSolutionsGreaterThanFourLowerChiSquarePValue}{p<.001}
\newcommand{\percentageFiveOrMoreCorrectTasksCtrlHigher}{29}
\newcommand{\percentageFiveOrMoreCorrectTasksTrmtHigher}{39}
\newcommand{\numberOfCorrectSolutionsGreaterThanFourHigherEffectSize}{1.52}
\newcommand{\numberOfCorrectSolutionsGreaterThanFourHigherChiSquarePValue}{p=.28}
\newcommand{\numberOfCorrectSolutionsGreaterThanFiveLowerEffectSize}{41.98}
\newcommand{\numberOfCorrectSolutionsGreaterThanFiveLowerChiSquarePValue}{p<.001}
\newcommand{\numberOfCorrectSolutionsGreaterThanFiveHigherEffectSize}{1.48}
\newcommand{\numberOfCorrectSolutionsGreaterThanFiveHigherChiSquarePValue}{p=.45}
\newcommand{\numberOfCorrectSolutionsGreaterThanSixLowerEffectSize}{22.94}
\newcommand{\numberOfCorrectSolutionsGreaterThanSixLowerChiSquarePValue}{p<.001}
\newcommand{\numberOfCorrectSolutionsGreaterThanSixHigherEffectSize}{8.67}
\newcommand{\numberOfCorrectSolutionsGreaterThanSixHigherChiSquarePValue}{p=.15}
\newcommand{\toolEvaluationTaskICohensKappa}{.63}
\newcommand{\toolEvaluationTaskIKendallsTau}{.68}
\newcommand{\toolEvaluationTaskIICohensKappa}{.76}
\newcommand{\toolEvaluationTaskIIKendallsTau}{.78}
\newcommand{\toolEvaluationTaskIIICohensKappa}{.76}
\newcommand{\toolEvaluationTaskIIIKendallsTau}{.78}
\newcommand{\toolEvaluationTaskIVCohensKappa}{1.00}
\newcommand{\toolEvaluationTaskIVKendallsTau}{1.00}
\newcommand{\toolEvaluationTaskVCohensKappa}{1.00}
\newcommand{\toolEvaluationTaskVKendallsTau}{1.00}
\newcommand{\toolEvaluationTaskVICohensKappa}{1.00}
\newcommand{\toolEvaluationTaskVIKendallsTau}{1.00}
\newcommand{\toolEvaluationTaskVIICohensKappa}{1.00}
\newcommand{\toolEvaluationTaskVIIKendallsTau}{1.00}
\newcommand{\toolEvaluationTaskVIIICohensKappa}{.75}
\newcommand{\toolEvaluationTaskVIIIKendallsTau}{.78}
\newcommand{\toolEvaluationTaskIXCohensKappa}{1.00}
\newcommand{\toolEvaluationTaskIXKendallsTau}{1.00}
\newcommand{\toolEvaluationTaskXCohensKappa}{1.00}
\newcommand{\toolEvaluationTaskXKendallsTau}{1.00}
\newcommand{\toolEvaluationTaskXICohensKappa}{1.00}
\newcommand{\toolEvaluationTaskXIKendallsTau}{1.00}
\newcommand{\toolEvaluationTaskXIICohensKappa}{1.00}
\newcommand{\toolEvaluationTaskXIIKendallsTau}{1.00}
\newcommand{\toolEvaluationTaskXIIICohensKappa}{1.00}
\newcommand{\toolEvaluationTaskXIIIKendallsTau}{1.00}
\newcommand{\toolEvaluationTaskXIVCohensKappa}{1.00}
\newcommand{\toolEvaluationTaskXIVKendallsTau}{1.00}
\newcommand{\toolEvaluationCompleteCohensKappa}{.96}
\newcommand{\toolEvaluationCompleteKendallsTau}{.96}
\newcommand{\taskIEffectSize}{2.20}
\newcommand{\taskIChiSquarePValue}{p=.051}
\newcommand{\taskIExperienceEffectSize}{2.35}
\newcommand{\taskIExperienceChiSquarePValue}{p=.096}
\newcommand{\taskINoExperienceEffectSize}{1.89}
\newcommand{\taskINoExperienceChiSquarePValue}{p=.51}
\newcommand{\taskIFemaleEffectSize}{1.69}
\newcommand{\taskIFemaleChiSquarePValue}{p=.25}
\newcommand{\taskIIEffectSize}{11.73}
\newcommand{\taskIIChiSquarePValue}{p<.001}
\newcommand{\taskIIExperienceEffectSize}{16.45}
\newcommand{\taskIIExperienceChiSquarePValue}{p<.001}
\newcommand{\taskIINoExperienceEffectSize}{8.98}
\newcommand{\taskIINoExperienceChiSquarePValue}{p=.002}
\newcommand{\taskIIFemaleEffectSize}{11.09}
\newcommand{\taskIIFemaleChiSquarePValue}{p<.001}
\newcommand{\taskIIIEffectSize}{2.26}
\newcommand{\taskIIIChiSquarePValue}{p=.018}
\newcommand{\taskIIIExperienceEffectSize}{1.75}
\newcommand{\taskIIIExperienceChiSquarePValue}{p=.24}
\newcommand{\taskIIINoExperienceEffectSize}{3.82}
\newcommand{\taskIIINoExperienceChiSquarePValue}{p=.027}
\newcommand{\taskIIIFemaleEffectSize}{2.24}
\newcommand{\taskIIIFemaleChiSquarePValue}{p=.031}
\newcommand{\taskIVEffectSize}{44.38}
\newcommand{\taskIVChiSquarePValue}{p<.001}
\newcommand{\taskIVExperienceEffectSize}{91.88}
\newcommand{\taskIVExperienceChiSquarePValue}{p<.001}
\newcommand{\taskIVNoExperienceEffectSize}{27.50}
\newcommand{\taskIVNoExperienceChiSquarePValue}{p<.001}
\newcommand{\taskIVFemaleEffectSize}{35.19}
\newcommand{\taskIVFemaleChiSquarePValue}{p<.001}
\newcommand{\taskVEffectSize}{.49}
\newcommand{\taskVChiSquarePValue}{p=.15}
\newcommand{\taskVExperienceEffectSize}{.21}
\newcommand{\taskVExperienceChiSquarePValue}{p=.031}
\newcommand{\taskVNoExperienceEffectSize}{1.24}
\newcommand{\taskVNoExperienceChiSquarePValue}{p>.99}
\newcommand{\taskVFemaleEffectSize}{.47}
\newcommand{\taskVFemaleChiSquarePValue}{p=.16}
\newcommand{\taskVIEffectSize}{8.31}
\newcommand{\taskVIChiSquarePValue}{p<.001}
\newcommand{\taskVIExperienceEffectSize}{13.78}
\newcommand{\taskVIExperienceChiSquarePValue}{p<.001}
\newcommand{\taskVINoExperienceEffectSize}{4.67}
\newcommand{\taskVINoExperienceChiSquarePValue}{p=.016}
\newcommand{\taskVIFemaleEffectSize}{9.24}
\newcommand{\taskVIFemaleChiSquarePValue}{p<.001}
\newcommand{\taskVIIEffectSize}{22.00}
\newcommand{\taskVIIChiSquarePValue}{p<.001}
\newcommand{\taskVIIExperienceEffectSize}{20.04}
\newcommand{\taskVIIExperienceChiSquarePValue}{p<.001}
\newcommand{\taskVIINoExperienceEffectSize}{37.50}
\newcommand{\taskVIINoExperienceChiSquarePValue}{p<.001}
\newcommand{\taskVIIFemaleEffectSize}{22.81}
\newcommand{\taskVIIFemaleChiSquarePValue}{p<.001}
\newcommand{\tasksLowerEffectSizeMean}{13.05}
\newcommand{\tasksLowerExperienceEffectSizeMean}{20.92}
\newcommand{\tasksLowerNoExperienceEffectSizeMean}{12.23}
\newcommand{\taskVIIIEffectSize}{1.53}
\newcommand{\taskVIIIChiSquarePValue}{p=.30}
\newcommand{\taskVIIIExperienceEffectSize}{1.05}
\newcommand{\taskVIIIExperienceChiSquarePValue}{p>.99}
\newcommand{\taskVIIINoExperienceEffectSize}{2.79}
\newcommand{\taskVIIINoExperienceChiSquarePValue}{p=.11}
\newcommand{\taskVIIIFemaleEffectSize}{1.52}
\newcommand{\taskVIIIFemaleChiSquarePValue}{p=.34}
\newcommand{\taskIXEffectSize}{1.13}
\newcommand{\taskIXChiSquarePValue}{p=.84}
\newcommand{\taskIXExperienceEffectSize}{.64}
\newcommand{\taskIXExperienceChiSquarePValue}{p=.44}
\newcommand{\taskIXNoExperienceEffectSize}{2.75}
\newcommand{\taskIXNoExperienceChiSquarePValue}{p=.11}
\newcommand{\taskIXFemaleEffectSize}{1.16}
\newcommand{\taskIXFemaleChiSquarePValue}{p=.80}
\newcommand{\taskXEffectSize}{1.63}
\newcommand{\taskXChiSquarePValue}{p=.17}
\newcommand{\taskXExperienceEffectSize}{2.36}
\newcommand{\taskXExperienceChiSquarePValue}{p=.051}
\newcommand{\taskXNoExperienceEffectSize}{.91}
\newcommand{\taskXNoExperienceChiSquarePValue}{p>.99}
\newcommand{\taskXFemaleEffectSize}{1.51}
\newcommand{\taskXFemaleChiSquarePValue}{p=.29}
\newcommand{\taskXIEffectSize}{1.60}
\newcommand{\taskXIChiSquarePValue}{p=.30}
\newcommand{\taskXIExperienceEffectSize}{1.94}
\newcommand{\taskXIExperienceChiSquarePValue}{p=.22}
\newcommand{\taskXINoExperienceEffectSize}{1.06}
\newcommand{\taskXINoExperienceChiSquarePValue}{p>.99}
\newcommand{\taskXIFemaleEffectSize}{1.72}
\newcommand{\taskXIFemaleChiSquarePValue}{p=.33}
\newcommand{\taskXIIEffectSize}{.88}
\newcommand{\taskXIIChiSquarePValue}{p=.82}
\newcommand{\taskXIIExperienceEffectSize}{.67}
\newcommand{\taskXIIExperienceChiSquarePValue}{p=.45}
\newcommand{\taskXIINoExperienceEffectSize}{1.42}
\newcommand{\taskXIINoExperienceChiSquarePValue}{p=.72}
\newcommand{\taskXIIFemaleEffectSize}{1.04}
\newcommand{\taskXIIFemaleChiSquarePValue}{p>.99}
\newcommand{\taskXIIIEffectSize}{1.04}
\newcommand{\taskXIIIChiSquarePValue}{p>.99}
\newcommand{\taskXIIIExperienceEffectSize}{.85}
\newcommand{\taskXIIIExperienceChiSquarePValue}{p=.90}
\newcommand{\taskXIIINoExperienceEffectSize}{1.52}
\newcommand{\taskXIIINoExperienceChiSquarePValue}{p=.73}
\newcommand{\taskXIIIFemaleEffectSize}{.92}
\newcommand{\taskXIIIFemaleChiSquarePValue}{p>.99}
\newcommand{\taskXIVEffectSize}{2.13}
\newcommand{\taskXIVChiSquarePValue}{p=.038}
\newcommand{\taskXIVExperienceEffectSize}{1.93}
\newcommand{\taskXIVExperienceChiSquarePValue}{p=.15}
\newcommand{\taskXIVNoExperienceEffectSize}{3.64}
\newcommand{\taskXIVNoExperienceChiSquarePValue}{p=.12}
\newcommand{\taskXIVFemaleEffectSize}{1.99}
\newcommand{\taskXIVFemaleChiSquarePValue}{p=.10}
\newcommand{\tasksHigherEffectSizeMean}{1.42}
\newcommand{\tasksHigherExperienceEffectSizeMean}{1.35}
\newcommand{\tasksHigherNoExperienceEffectSizeMean}{2.01}
\newcommand{\processingTimeTaskIWilcoxPValue}{p=.044}
\newcommand{\processingTimeTaskIEffectSize}{.59}
\newcommand{\processingTimeJustCorrectTaskIWilcoxPValue}{p=.68}
\newcommand{\processingTimeJustCorrectTaskIEffectSize}{.52}
\newcommand{\processingTimeJustWrongTaskIWilcoxPValue}{p=.025}
\newcommand{\processingTimeJustWrongTaskIEffectSize}{.72}
\newcommand{\processingTimeExperienceTaskIWilcoxPValue}{p=.050}
\newcommand{\processingTimeExperienceTaskIEffectSize}{.61}
\newcommand{\processingTimeNoExperienceTaskIWilcoxPValue}{p=.29}
\newcommand{\processingTimeNoExperienceTaskIEffectSize}{.58}
\newcommand{\processingTimeTaskIIWilcoxPValue}{p<.001}
\newcommand{\processingTimeTaskIIEffectSize}{.86}
\newcommand{\processingTimeJustCorrectTaskIIWilcoxPValue}{p<.001}
\newcommand{\processingTimeJustCorrectTaskIIEffectSize}{.82}
\newcommand{\processingTimeJustWrongTaskIIWilcoxPValue}{p=.022}
\newcommand{\processingTimeJustWrongTaskIIEffectSize}{.82}
\newcommand{\processingTimeExperienceTaskIIWilcoxPValue}{p<.001}
\newcommand{\processingTimeExperienceTaskIIEffectSize}{.84}
\newcommand{\processingTimeNoExperienceTaskIIWilcoxPValue}{p<.001}
\newcommand{\processingTimeNoExperienceTaskIIEffectSize}{.91}
\newcommand{\processingTimeTaskIIIWilcoxPValue}{p<.001}
\newcommand{\processingTimeTaskIIIEffectSize}{.72}
\newcommand{\processingTimeJustCorrectTaskIIIWilcoxPValue}{p<.001}
\newcommand{\processingTimeJustCorrectTaskIIIEffectSize}{.73}
\newcommand{\processingTimeJustWrongTaskIIIWilcoxPValue}{p=.011}
\newcommand{\processingTimeJustWrongTaskIIIEffectSize}{.69}
\newcommand{\processingTimeExperienceTaskIIIWilcoxPValue}{p<.001}
\newcommand{\processingTimeExperienceTaskIIIEffectSize}{.76}
\newcommand{\processingTimeNoExperienceTaskIIIWilcoxPValue}{p=.018}
\newcommand{\processingTimeNoExperienceTaskIIIEffectSize}{.68}
\newcommand{\processingTimeTaskIVWilcoxPValue}{p<.001}
\newcommand{\processingTimeTaskIVEffectSize}{.81}
\newcommand{\processingTimeJustCorrectTaskIVWilcoxPValue}{p=.34}
\newcommand{\processingTimeJustCorrectTaskIVEffectSize}{.60}
\newcommand{\processingTimeJustWrongTaskIVWilcoxPValue}{p=.006}
\newcommand{\processingTimeJustWrongTaskIVEffectSize}{.73}
\newcommand{\processingTimeExperienceTaskIVWilcoxPValue}{p<.001}
\newcommand{\processingTimeExperienceTaskIVEffectSize}{.83}
\newcommand{\processingTimeNoExperienceTaskIVWilcoxPValue}{p<.001}
\newcommand{\processingTimeNoExperienceTaskIVEffectSize}{.77}
\newcommand{\processingTimeTaskVWilcoxPValue}{p=.74}
\newcommand{\processingTimeTaskVEffectSize}{.48}
\newcommand{\processingTimeJustCorrectTaskVWilcoxPValue}{p=.58}
\newcommand{\processingTimeJustCorrectTaskVEffectSize}{.47}
\newcommand{\processingTimeJustWrongTaskVWilcoxPValue}{p=.067}
\newcommand{\processingTimeJustWrongTaskVEffectSize}{.72}
\newcommand{\processingTimeExperienceTaskVWilcoxPValue}{p=.93}
\newcommand{\processingTimeExperienceTaskVEffectSize}{.50}
\newcommand{\processingTimeNoExperienceTaskVWilcoxPValue}{p=.52}
\newcommand{\processingTimeNoExperienceTaskVEffectSize}{.45}
\newcommand{\processingTimeTaskVIWilcoxPValue}{p<.001}
\newcommand{\processingTimeTaskVIEffectSize}{.82}
\newcommand{\processingTimeJustCorrectTaskVIWilcoxPValue}{p<.001}
\newcommand{\processingTimeJustCorrectTaskVIEffectSize}{.73}
\newcommand{\processingTimeJustWrongTaskVIWilcoxPValue}{p=.004}
\newcommand{\processingTimeJustWrongTaskVIEffectSize}{.79}
\newcommand{\processingTimeExperienceTaskVIWilcoxPValue}{p<.001}
\newcommand{\processingTimeExperienceTaskVIEffectSize}{.79}
\newcommand{\processingTimeNoExperienceTaskVIWilcoxPValue}{p<.001}
\newcommand{\processingTimeNoExperienceTaskVIEffectSize}{.87}
\newcommand{\processingTimeTaskVIIWilcoxPValue}{p=.004}
\newcommand{\processingTimeTaskVIIEffectSize}{.63}
\newcommand{\processingTimeJustCorrectTaskVIIWilcoxPValue}{p=.13}
\newcommand{\processingTimeJustCorrectTaskVIIEffectSize}{.64}
\newcommand{\processingTimeJustWrongTaskVIIWilcoxPValue}{p=.15}
\newcommand{\processingTimeJustWrongTaskVIIEffectSize}{.62}
\newcommand{\processingTimeExperienceTaskVIIWilcoxPValue}{p=.013}
\newcommand{\processingTimeExperienceTaskVIIEffectSize}{.64}
\newcommand{\processingTimeNoExperienceTaskVIIWilcoxPValue}{p=.14}
\newcommand{\processingTimeNoExperienceTaskVIIEffectSize}{.61}
\newcommand{\processingTimeLowerEffectSizeMean}{.70}
\newcommand{\processingTimeLowerJustCorrectEffectSizeMean}{.65}
\newcommand{\processingTimeLowerJustWrongEffectSizeMean}{.73}
\newcommand{\processingTimeLowerExperienceEffectSizeMean}{.71}
\newcommand{\processingTimeLowerNoExperienceEffectSizeMean}{.70}
\newcommand{\processingTimeTaskVIIIWilcoxPValue}{p=.029}
\newcommand{\processingTimeTaskVIIIEffectSize}{.60}
\newcommand{\processingTimeJustCorrectTaskVIIIWilcoxPValue}{p=.58}
\newcommand{\processingTimeJustCorrectTaskVIIIEffectSize}{.53}
\newcommand{\processingTimeJustWrongTaskVIIIWilcoxPValue}{p=.029}
\newcommand{\processingTimeJustWrongTaskVIIIEffectSize}{.69}
\newcommand{\processingTimeExperienceTaskVIIIWilcoxPValue}{p=.43}
\newcommand{\processingTimeExperienceTaskVIIIEffectSize}{.55}
\newcommand{\processingTimeNoExperienceTaskVIIIWilcoxPValue}{p=.011}
\newcommand{\processingTimeNoExperienceTaskVIIIEffectSize}{.69}
\newcommand{\processingTimeTaskIXWilcoxPValue}{p=.065}
\newcommand{\processingTimeTaskIXEffectSize}{.58}
\newcommand{\processingTimeJustCorrectTaskIXWilcoxPValue}{p=.007}
\newcommand{\processingTimeJustCorrectTaskIXEffectSize}{.65}
\newcommand{\processingTimeJustWrongTaskIXWilcoxPValue}{p=.51}
\newcommand{\processingTimeJustWrongTaskIXEffectSize}{.45}
\newcommand{\processingTimeExperienceTaskIXWilcoxPValue}{p=.51}
\newcommand{\processingTimeExperienceTaskIXEffectSize}{.54}
\newcommand{\processingTimeNoExperienceTaskIXWilcoxPValue}{p=.034}
\newcommand{\processingTimeNoExperienceTaskIXEffectSize}{.66}
\newcommand{\processingTimeTaskXWilcoxPValue}{p=.022}
\newcommand{\processingTimeTaskXEffectSize}{.60}
\newcommand{\processingTimeJustCorrectTaskXWilcoxPValue}{p=.43}
\newcommand{\processingTimeJustCorrectTaskXEffectSize}{.55}
\newcommand{\processingTimeJustWrongTaskXWilcoxPValue}{p=.22}
\newcommand{\processingTimeJustWrongTaskXEffectSize}{.58}
\newcommand{\processingTimeExperienceTaskXWilcoxPValue}{p=.007}
\newcommand{\processingTimeExperienceTaskXEffectSize}{.66}
\newcommand{\processingTimeNoExperienceTaskXWilcoxPValue}{p>.99}
\newcommand{\processingTimeNoExperienceTaskXEffectSize}{.50}
\newcommand{\processingTimeTaskXIWilcoxPValue}{p=.86}
\newcommand{\processingTimeTaskXIEffectSize}{.51}
\newcommand{\processingTimeJustCorrectTaskXIWilcoxPValue}{p=.89}
\newcommand{\processingTimeJustCorrectTaskXIEffectSize}{.52}
\newcommand{\processingTimeJustWrongTaskXIWilcoxPValue}{p=.82}
\newcommand{\processingTimeJustWrongTaskXIEffectSize}{.49}
\newcommand{\processingTimeExperienceTaskXIWilcoxPValue}{p=.15}
\newcommand{\processingTimeExperienceTaskXIEffectSize}{.58}
\newcommand{\processingTimeNoExperienceTaskXIWilcoxPValue}{p=.12}
\newcommand{\processingTimeNoExperienceTaskXIEffectSize}{.38}
\newcommand{\processingTimeTaskXIIWilcoxPValue}{p=.94}
\newcommand{\processingTimeTaskXIIEffectSize}{.50}
\newcommand{\processingTimeJustCorrectTaskXIIWilcoxPValue}{p=.14}
\newcommand{\processingTimeJustCorrectTaskXIIEffectSize}{.61}
\newcommand{\processingTimeJustWrongTaskXIIWilcoxPValue}{p=.26}
\newcommand{\processingTimeJustWrongTaskXIIEffectSize}{.44}
\newcommand{\processingTimeExperienceTaskXIIWilcoxPValue}{p=.36}
\newcommand{\processingTimeExperienceTaskXIIEffectSize}{.45}
\newcommand{\processingTimeNoExperienceTaskXIIWilcoxPValue}{p=.22}
\newcommand{\processingTimeNoExperienceTaskXIIEffectSize}{.59}
\newcommand{\processingTimeTaskXIIIWilcoxPValue}{p=.70}
\newcommand{\processingTimeTaskXIIIEffectSize}{.52}
\newcommand{\processingTimeJustCorrectTaskXIIIWilcoxPValue}{p=.92}
\newcommand{\processingTimeJustCorrectTaskXIIIEffectSize}{.49}
\newcommand{\processingTimeJustWrongTaskXIIIWilcoxPValue}{p=.96}
\newcommand{\processingTimeJustWrongTaskXIIIEffectSize}{.49}
\newcommand{\processingTimeExperienceTaskXIIIWilcoxPValue}{p=.93}
\newcommand{\processingTimeExperienceTaskXIIIEffectSize}{.51}
\newcommand{\processingTimeNoExperienceTaskXIIIWilcoxPValue}{p=.49}
\newcommand{\processingTimeNoExperienceTaskXIIIEffectSize}{.55}
\newcommand{\processingTimeTaskXIVWilcoxPValue}{p=.12}
\newcommand{\processingTimeTaskXIVEffectSize}{.57}
\newcommand{\processingTimeJustCorrectTaskXIVWilcoxPValue}{p=.81}
\newcommand{\processingTimeJustCorrectTaskXIVEffectSize}{.52}
\newcommand{\processingTimeJustWrongTaskXIVWilcoxPValue}{p=.50}
\newcommand{\processingTimeJustWrongTaskXIVEffectSize}{.54}
\newcommand{\processingTimeExperienceTaskXIVWilcoxPValue}{p=.47}
\newcommand{\processingTimeExperienceTaskXIVEffectSize}{.54}
\newcommand{\processingTimeNoExperienceTaskXIVWilcoxPValue}{p=.11}
\newcommand{\processingTimeNoExperienceTaskXIVEffectSize}{.62}
\newcommand{\processingTimeHigherEffectSizeMean}{.55}
\newcommand{\processingTimeHigherJustCorrectEffectSizeMean}{.55}
\newcommand{\processingTimeHigherJustWrongEffectSizeMean}{.52}
\newcommand{\processingTimeHigherExperienceEffectSizeMean}{.54}
\newcommand{\processingTimeHigherNoExperienceEffectSizeMean}{.57}
\newcommand{\processingTimeLowerWilcoxPValue}{p<.001}
\newcommand{\processingTimeLowerEffectSize}{.71}
\newcommand{\processingTimeHigherWilcoxPValue}{p=.002}
\newcommand{\processingTimeHigherEffectSize}{.55}
\newcommand{\processingTimeJustCorrectLowerWilcoxPValue}{p<.001}
\newcommand{\processingTimeJustCorrectLowerEffectSize}{.62}
\newcommand{\processingTimeJustCorrectHigherWilcoxPValue}{p=.052}
\newcommand{\processingTimeJustCorrectHigherEffectSize}{.55}
\newcommand{\processingTimeJustWrongLowerWilcoxPValue}{p<.001}
\newcommand{\processingTimeJustWrongLowerEffectSize}{.72}
\newcommand{\processingTimeJustWrongHigherWilcoxPValue}{p=.62}
\newcommand{\processingTimeJustWrongHigherEffectSize}{.51}
\newcommand{\processingTimeExperienceLowerWilcoxPValue}{p<.001}
\newcommand{\processingTimeExperienceLowerEffectSize}{.71}
\newcommand{\processingTimeExperienceHigherWilcoxPValue}{p=.049}
\newcommand{\processingTimeExperienceHigherEffectSize}{.54}
\newcommand{\processingTimeNoExperienceLowerWilcoxPValue}{p<.001}
\newcommand{\processingTimeNoExperienceLowerEffectSize}{.70}
\newcommand{\processingTimeNoExperienceHigherWilcoxPValue}{p=.014}
\newcommand{\processingTimeNoExperienceHigherEffectSize}{.57}

%% file: content/introduction.tex
\section{Introduction}
\label{sec:introduction}

% Context: Learners, their misconceptions, which manifest in bugs
Block-based programming is frequently used as entry-point to computational thinking and to programming, since the block-based nature reduces complexity compared to text-based programming languages. Nevertheless, there are many concepts to comprehend, and misconceptions about these concepts may hamper progress in learning. These misconceptions often manifest in ``buggy'' programs, i.e., programs that contain mistakes and do not function correctly. It is then up to the teacher to help learners identify these bugs in their programs, and to fix or explain them in order to overcome the misconceptions that caused them. 
% Problem: Teachers need to possess the skills to identify and correct these bugs in student programs.
% shortened: While debugging an individual program can already be challenging, the classroom setting exacerbates this problem: Teachers may face many buggy programs at the same time, and the common practice of using open-ended programming tasks means that these may be creative programs the teacher has never seen before.
While subject teachers at secondary schools may be expected to be adequately educated in debugging, programming is increasingly also introduced at primary schools, where teachers are not specialised in all individual subjects they teach and thus may not have adequate training.

% teachers may be faced with many buggy programs at the same time in a classroom setting. The common practice of using open-ended programming tasks, where learners develop creative programs the teacher has never seen before, further exacerbates the problem.
% %
% While the
% %
% %Since block-based programming is increasingly taught at early stages of education, this means that even primary school teachers require strong debugging skills.
% While in secondary schools, teachers are mostly subject teachers, primary school teachers teach nearly all subjects. Thus, primary school teachers should also acquire background knowledge about programming concepts in their teacher training.

\begin{figure}[t]
	\centering
	\subfloat[\label{fig:mls_bug}Buggy version.]{\includegraphics[width=0.45\columnwidth]{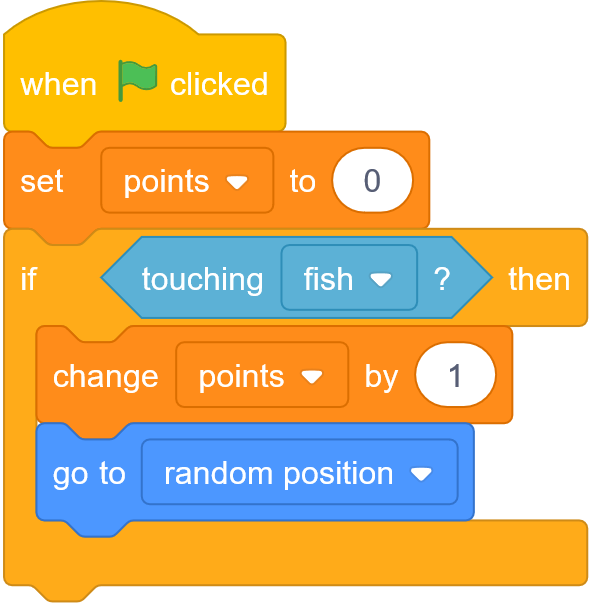}}\hspace{2em}
	\subfloat[\label{fig:mls_fix}Correct version.]{\includegraphics[width=0.45\columnwidth]{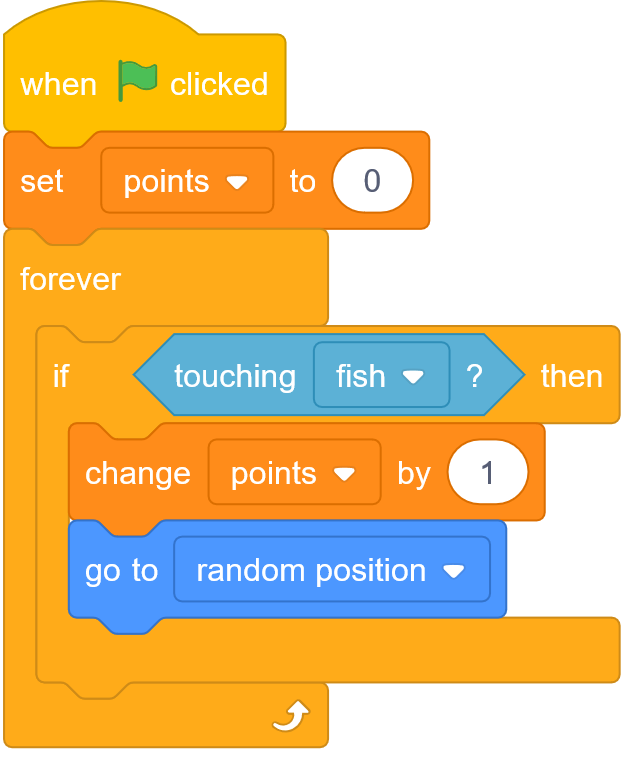}}
	\caption{An example of the \textit{Missing Loop Sensing} bug pattern: \textmd{Instead of continuously checking for collisions with a \emph{fish} sprite, the buggy snippet performs just a single check immediately after initialising the variable \emph{points}---this will scarcely suffice to detect all possible touching events.}} 
		%Without the forever-loop, the check whether the `fish' sprite is touched is only performed once, immediately after initialising the points, rather than continuously.}
	\label{fig:mlsexample}
%	\vspace{-1.5em}
\end{figure}

% Insight: Some bugs are so common that we can describe them by common bug patterns. We can even identify some of them automatically (LitterBox).
Teachers have the advantage that learners frequently have similar misconceptions, which result in recurring patterns of bugs in their programs. 
%
% Motivational example
For example, a common misconception of learners is that an \emph{if}-statement triggers each time the condition becomes true~\cite{sorva2018}. \Cref{fig:mls_bug}
shows a code snippet from a \scratch~\cite{maloney2010} program where this misconception manifests in a bug. %: The script checks for the occurrence of a touch-event only once, which means that the enclosed code is only executed if the condition is fulfilled immediately after initialising the variable \emph{points}. 
A correct implementation (\Cref{fig:mls_fix}) would enclose the check inside a \emph{forever}-loop to continuously check for the occurrence of the event. Bugs matching this pattern occur frequently: In a recent study, this \emph{Missing Loop Sensing} bug pattern was found in 3,282 projects in a random dataset of
%135,164 
74,830 \scratch projects~\cite{fraedrich2020}. 

% Contribution
%Bug patterns provide an opportunity to support teachers 
%%If teachers have knowledge of the bug patterns that occur in their students' %programs, they may be more efficient and more effective at identifying the bugs %and the misconceptions that caused them. Teachers can even be supported with 
%by identifying them automatically and providing hints on them. 

Instances of bug patterns like the \emph{Missing Loop Sensing} example above can be detected automatically. The repetitive nature of these bug patterns also makes it possible to provide generic hints on how to fix the identified bugs, and what misconceptions may cause them. %(\cref{fig:hint_example}).
However, especially primary school students might be overwhelmed by the concepts and the technical terminology used in automatically generated hints. Therefore it might be advisable that the teacher deals with the hints and not the primary school student. The teachers can then consider which information they communicate depending, e.g., on the student's skills.
While one would expect that an appropriate hint pointing out the location and type of a bug simplifies debugging, the same information may hamper the learning of teachers \emph{in training}: The use of direct instruction has been reported to have negative effects on transferring compared to using a more discovery based learning approach~\cite{kapur2012designing}. 
Intuitively, the act of manually debugging a program is a valuable learning experience, and reliance on a tool may inhibit the teachers' ability to spot similar bugs on their own, when no such hints are provided.
In order to study the effects of hints on bug patterns, in this paper we therefore aim to investigate these two aspects:
\begin{itemize}
	\item How do hints on bug patterns influence the effectiveness and
efficiency of teachers while debugging and fixing bugs? In order to support
research on bug patterns and their detection, and to provide evidence for teachers to decide whether they should apply bug detection tools, we aim to determine empirically what the benefits of showing hints on bug patterns are for teachers.
	\item How do hints on bug patterns influence the ability of teachers in training to recognise and fix bugs without hints? In order to support teacher training, we aim to determine empirically whether providing detailed hints on bug patterns during training affects whether the teachers in training can fix similar bugs when not given any hints.
\end{itemize}

To answer these questions, we conducted an experiment with
\numberOfParticipants~primary school teachers in training. We first compared
the performance in debugging and fixing broken \scratch programs between a subgroup of our participants who were given hints on bug patterns (\groupB) versus a subgroup not receiving any hints (\groupA). This allows us to quantify effects on their success in fixing the bugs as well as the time it took them to do so. 
In order to evaluate which learning experiences can be transferred better, we then experimentally determined whether the participants of \groupA were able to spot similar bugs in other programs more easily without the hints than \groupB.
This provides us with information whether hints can be transferred, or whether missing the learning experience of manually identifying the bugs hampers the debugging skills of teachers in training.

% Outlook to results
Our experiment confirms that giving hints on bug patterns improves the performance at correcting bugs described by the hints: In general, hints can reduce the effort of finding and fixing bugs in terms of time from \meanProcessingTimeCtrlLower~to \meanProcessingTimeTrmtLower~minutes, while increasing the effectiveness in terms of correct programs by \percentageFunctionalityDifferenceLower\% more correct solutions. This suggests that providing tool-support to recognise bug patterns is desirable for teachers in practice, and research on identifying bug patterns has the potential to improve classroom learning.
We do, however, observe that the beneficial effects are larger for participants having prior programming experience. Although our experiment provides no significant evidence that either learning to debug with hints or learning to debug ``the hard way'' leads to better learning effects, the knowledge of the bug patterns that was gained by the hints seems to slightly improve performance on similar tasks without hints. This suggests that bug patterns might be a useful concept to include in the curriculum for teachers in training.

%% file: content/relatedwork.tex
\section{Related Work}
\label{sec:relatedwork}

\begin{figure}[t!]
	\centering
	\includegraphics[width=1\columnwidth]{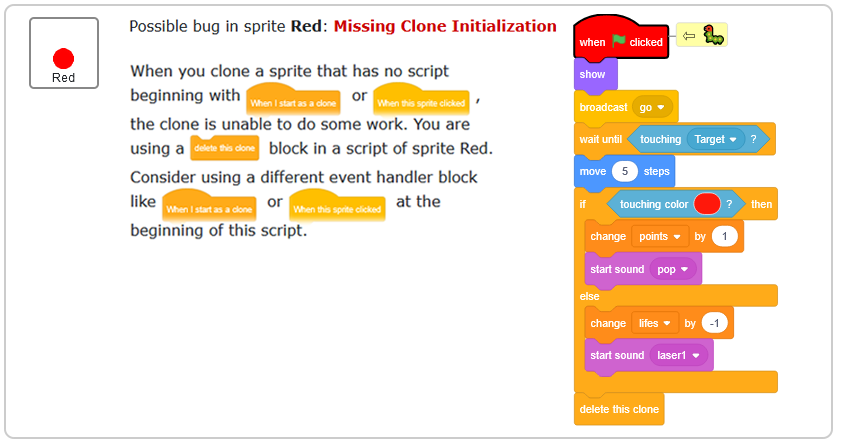}
	\vspace{-1em}
	\caption{\label{fig:hint_example_MCI} Example hint provided by \litterbox to support recognising and fixing the \emph{Missing Clone Initialisation} bug pattern present in Task 4.}
	\vspace{-1em}
\end{figure}

%\begin{figure}[t!]
%	\centering
%	\includegraphics[width=1\columnwidth]{grafics/task02Hint.png}
%	\vspace{-1em}
%	\caption{\label{fig:hint_example} Example hint provided by \litterbox to support recognising and fixing the \emph{Missing Loop Sensing} bug present in Task 2.}
		%Hint for Task 2 on \emph{Missing Loop Sensing} as generated by \litterbox.}
%	\vspace{-1em}
%\end{figure}

\subsection{Misconceptions and Bug Patterns}
Learners of programming can have misconceptions about programming constructs \cite{swidan2018}. Such misconceptions may severely hamper their ability to progress, as they will lead to programs that do not work and are a source of frustration~\cite{hansen2007}.
In the context of beginner programming, Sorva~\cite{sorva2018} identified a catalogue of 41 such common misconceptions; for example, the misconception that variables can store multiple values and store the history of values assigned to it, or that a while loop's condition is evaluated continuously and the loop exits the instant it becomes false.
Common misconceptions may result in similarly erroneous (i.e., \emph{buggy}) programs. Even though programs may be functionally very different, the bugs resulting from misconceptions may be very similar. Fraedrich et al.~\cite{fraedrich2020} introduced the notion of common \emph{patterns} of such bugs. Generally speaking, a bug pattern in a block-based programming language is a composition of blocks that is typical of defective code~\cite{fraedrich2020}. Such bug patterns can be observed frequently in practice: Fraedrich et al.~\cite{fraedrich2020} found that out of a sample of 74,830 publicly shared \scratch projects, 33,655 contained at least one instance of a bug pattern. 

Of course, not all bugs result in bug patterns. As bug patterns do not deal with task-specific information but the code itself, bugs such as using the wrong variable, the wrong content in e.g. \emph{say} blocks or forgetting to implement certain features cannot be detected. Such content-related %logical
bugs can be addressed with dynamic analysis tools for \scratch such as \whisker \cite{stahlbauer2019testing}. They run the code and check if specified events occur or not. This  implies that the intended aim of the program must be known and respective tests must be available or created by the teacher. This is why dynamic testing cannot be easily applied to programs that result from open programming tasks and without feeding the tool with further individualised information, no feedback on how to proceed can be provided (apart from that the feature must be changed or implemented). 

In contrast to that, static analysis tools enable teachers and students to analyse programs without having to specify its intended output. Furthermore, when such tools check if instances of bug patterns exist they can provide generic information on the detected bug patterns, such as knowledge on the associated misconception or on how to remove the respective bug pattern. This is done by the \litterbox~\cite{fraser2021} tool as it can be seen in the example hint in \cref{fig:hint_example_MCI}.
Intuitively, such knowledge can aid in the process of determining whether code is correct, or in the process of debugging defective code: If an instance of a bug pattern has been spotted, this is a suitable point to start debugging. Indeed, bug patterns tend to lead to functionally defective programs: Out of 250 manually analysed projects 218 contained bug patterns that result in erroneous behaviour~\cite{fraedrich2020}.
%The \emph{Missing Loop Sensing} bug pattern, shown in \cref{fig:mlsexample} in the previous section, is an example of the 25 bug patterns introduced by Fraedrich et al.~\cite{fraedrich2020}. It is one of many possible misconceptions beginners tend to have about control flow in programs, when they confuse conditional statements and processing of events.
%
Other existing code analysis tools for \scratch such as \hairball~\cite{boe2013}, \qualityhound~\cite{techapaloku2017b} or \drscratch~\cite{moreno2015} report general quality problems in terms of code smells (= unaesthetic or less readable code that does not harm the functionality of a program), rather than specific bug patterns resulting from misconceptions. 
Previous research has shown that these code smells in \scratch hamper novices' learning processes~\cite{hermans2016b}. Bug patterns are more severe than code smells, thus likely having an even worse impact on novice programmers; our investigations in this paper suggest a possible way of dealing with them.

\subsection{Feedback on Code}
The ability to detect such bug patterns in student code makes it possible to provide learners and teachers with feedback.
Kennedy et al.~\cite{kennedy2020} demonstrated---using peer feedback and discussions of coding assignments---that misconceptions can be cleared with feedback in principle. In this paper, we evaluate the effects of feedback that is generated automatically by static analysis. 
Past research on automated hint generation has mainly considered the problem of providing hints on what should be the next step in solving programming assignments~\cite{singh2012,price2017b,zimmerman2015,rivers2015,wang2020} or open ended programming tasks~\cite{price2016,paassen2017} and how novices seek help in these systems~\cite{price2017c,marwan2020,aleven2016help,marwan2019evaluation}. 
An exception is the work by Gusukuma et al.~\cite{gusukuma2018}, who showed that feedback delivery on mistakes that anticipate possible misconceptions generally leads to favourable results, and that showing such hints does not harm transfer to new tasks. In contrast to this work, the hints we evaluate are not tailored to our self-study material and thus might be more universally applicable.

\subsection{Teachers as Debuggers}

% challenges
Not only students but also teachers---who are expected to support their students---struggle with debugging. Sentance and Csizmadia~\cite{sentance2017computing} examined the opinions of teachers in the UK about challenges when teaching computing. The most frequently mentioned challenge was the teacher's subject knowledge, and in particular this was mentioned more often by primary school teachers than by secondary school teachers. Within this challenge, teachers were concerned if they were able to help their students with their problems when programming.
Yadav et al.~\cite{yadav2016expanding} came to similar conclusions after interviewing high school computer science teachers, where they identified both content and pedagogical challenges. Supporting students when teaching programming is perceived as a main difficulty---not only because of the partially missing computer science background: The teachers explained this challenge also with (1)~the teacher-student ratio and (2)~the different programming approaches of the students that result in different needs.
% debugging performance
Teachers' debugging performance has been studied explicitly by Kim et al.~\cite{kim2018debugging}, who examined what kind of bugs early childhood preservice teachers produce and how they deal with given bugs: %The study detected six common bugs and showed that 
The teachers had difficulties debugging given programs, even though they tried different debugging strategies.
Michaeli and Romeike~\cite{michaeli2019current} interviewed high school teachers on their strategies to cope with programming bugs of their students. Most teachers reported that they have to help efficiently to reach all students with problems and that there is no time for detailed explanations. 
Consequently, there is a need to examine how teachers---especially those with insufficient subject knowledge---can be supported in debugging. The aim of this paper is to generate an initial understanding of the effects of hints on bug patterns. Do such hints have a positive influence on the effectiveness, efficiency and learning opportunities of teachers inspecting their students' programs?

%% file: content/study.tex
\section{Study Setup}
\label{sec:study}

\begin{figure}[tb]
	\centering
	\includegraphics[width=0.45\textwidth]{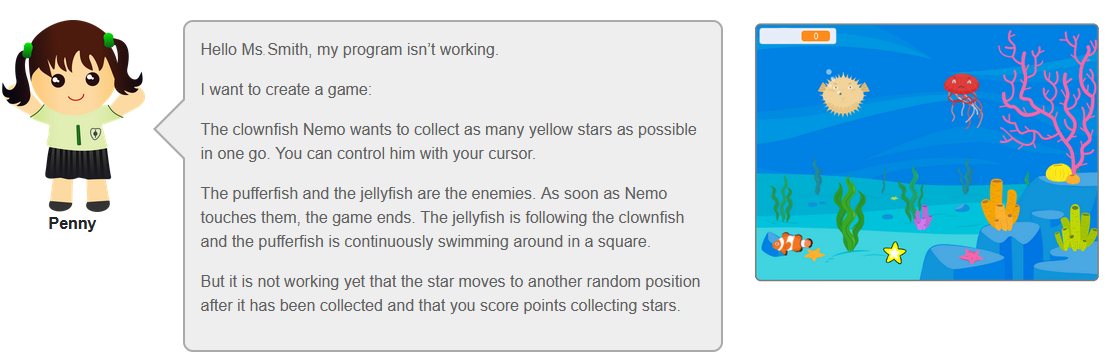}
%	\vspace{-1em}
	\caption{\label{fig:description_example} 
		Fictitious student addressing a study participant as part of Task 9.}
	\vspace{-1em}
\end{figure}

\begin{table*}[t]\centering
	\caption{Bug patterns and corresponding hints.}
	\label{tab:hints}
%	\vspace{-1em}
\begin{tabular}{p{2.5cm}p{15cm}}
\toprule
\textbf{Bug Pattern} & \textbf{\litterbox Hint} \\
\midrule
Message Never Sent (Tasks 1 and 8) & The message ``\emph{Game Over}'', that is to be received here, is never sent. Therefore, the adjacent script will never be triggered. \newline
If you want to receive a message, you have to select a message, that is already sent via a different script or you have to create and send a matching message inside a different script.\\
\midrule

Missing Loop Sensing (Tasks 2 and 9) & The highlighted condition is checked only once. Thus, the script runs through too fast. \newline
Implement a continuous check by surrounding the conditional statement with a \emph{forever loop}.\\
\midrule

Comparing Literals (Tasks 3 and 10) & You used \emph{current year} just as plain text. Your comparison will always return FALSE. Therefore, the code following \emph{wait until} is never executed. \newline
Did you rather intend to use your already existing variable block \emph{current year} instead of the plain text? You will find this block in the \emph{Variables} toolbox. \\
\midrule

Missing Clone Initialisation (Tasks 4 and 11) & When you clone a sprite that has no script beginning with \emph{When I start as a clone} or \emph{When this sprite clicked}, the clone is unable to do some work.\newline You are using a \emph{delete this clone} block in a script of \emph{Sprite Red}. Consider using a different event handler block like \emph{When I start as a clone} or \emph{When this sprite clicked} at the beginning of this script.\\
\midrule

Message Never Received (Tasks 5 and 12) & The message ``\emph{player touches money}'' that is sent here is never received by a \emph{when I receive ``player touches money''} block. Therefore nothing will happen as a reaction to this message. \newline
When you send a message make sure that another script receives it.\\
\midrule

Forever Inside Loop (Tasks 6 and 13) & The inner \emph{forever loop} is never left. Therefore, all blocks in this script but outside of the forever loop are never executed again. \newline
Try omitting the inner forever loop. \\
\midrule

Stuttering Movement (Tasks 7 and 14) & If you continuously press a key, you expect a smooth event processing. Unfortunately, a delay occurs between the first and the second processing round, resulting in stuttering movement. \newline
You can prevent this delay to happen by using the \emph{key right arrow pressed?} block from the \emph{Sensing} toolbox. To do that, you have to put the conditional statement \emph{if key right arrow pressed? then} inside of a \emph{forever loop} and use the event handler \emph{When green flag clicked} instead of the event handler \emph{When key right arrow pressed}.\\

\bottomrule
\end{tabular}
\end{table*}

To evaluate the impact of hints on bug patterns on the performance of teachers, we aim to answer the following research questions:

\textbf{RQ 1:} How do hints on bug patterns influence the effectiveness and efficiency of teachers while debugging and fixing bugs? %Do hints on bug patterns in failing programs increase the participants' performance on repairing bugs?

\textbf{RQ 2:} How do hints on bug patterns influence the ability of teachers in training to recognise and fix bugs without hints? %Do hints on bug patterns affect the participants' ability to correct similar bugs in other programs without hints?
	%

% Study Design
To answer these research questions, we conducted an A/B study with primary school teachers in training, who were tasked to fix several buggy \scratch programs, with and without hints.

\subsection{Study Participants}
\label{sec:participants}

% Participants
We implemented a three-week programming session into a course on mathematical didactics for primary schools teachers in training with \numLectureParticipants 
students signed up at the University of Passau. 
We chose the participants of this course because primary school teachers have to teach almost every subject but they are not specialised in all of them. Thus, they may not have had adequate training and may lack adequate skills for debugging. Although computer science is not yet part of the Bavarian curriculum, programming is increasingly introduced in primary schools. 

While course participants were recommended to take part in the programming sessions, it was not mandatory as they had a choice of which sessions of the overall course their assessment should be based on. Of the registered students, \numberOfActiveParticipants~actively participated, but when participants did not submit all tasks, submitted corrupted files, or submitted programs for the wrong tasks, we conservatively excluded all data of these participants.
Overall, we excluded all data points of \droupoutsTotal~participants. 
%By this, it is easier to detect differences between the tasks as there is the same amount of e.g. programs and hint evaluations for each task. 
We excluded \dropoutsCtrlFalseSubmissions~participants of \groupA and \dropoutsTrmtFalseSubmissions~participants of \groupB because some of their submitted programs were corrupted or because they submitted programs of previous tasks. Additionally, we excluded \dropoutsCtrlNotAllTasks~participants of \groupA and \dropoutsTrmtNotAllTasks~of \groupB because they did not submit all tasks. This allows us to compare the performance across tasks, which would be challenging if participants would differ between tasks. 

This results in a total of \numberOfParticipants~participants, of which \numberOfFemaleParticipants~were female, \numberOfMaleParticipants~were male, and they were between \ageYoungestParticipant~and \ageOldestParticipant~years of age with a median of \ageParticipantsMedian~years. According to our demographic pre-survey, many participants had no prior experience in programming at all, which means they neither programmed at school nor at university or anywhere else (\groupA: \numberOfUnexeriencedParticipantsCtrl\%; \groupB: \numberOfUnexeriencedParticipantsTrmt\%). We will refer to these participants as \emph{inexperienced} participants and to those with some prior  experience as \emph{experienced} participants. 

To ensure all participants had sufficient basic knowledge for the experiment, we provided self-study material that can be completed in 90 minutes: Within one week, the participants were tasked to watch a 30 minutes explanatory video and perform three exercises. In the video a \scratch program is built step by step, thereby covering all blocks and concepts that were needed to fix the buggy \scratch programs later in the experiment. Specific misconceptions and bug patterns were not discussed. The participants had to stop the video three times to perform an exercise. This approach is based on the Use-Modify-Create-Framework~\cite{lee2011computational}: At first, they used the presented program by opening, running and interpreting it. In the second exercise, they expanded the functionality of one sprite. Finally, they created their own program in about 30 minutes.

\subsection{Experiment Procedure}
\label{sec:procedure}

\subsubsection{Overview}

After introducing the participants to \scratch with the self-study material (described in \cref{sec:participants}), the main part of the experiment started: The participants were tasked to fix one example program (Task 0), which served to familiarise participants with the infrastructure, and 14 buggy \scratch programs (Tasks 1--14) within two weeks using their own digital devices. %All 15 buggy programs are available at \emph{blinded}.
They should spend about ten minutes on one task but were allowed to go on to the next task, regardless of the current task being completed or not. However, once they had proceeded to the next task, they were not able to go back to previous tasks. The tasks were provided on a website created by the researchers. The participants downloaded the broken programs, edited them on the public \scratch website and then again uploaded the edited programs to our website---even if they could not fix it. The participants were asked to pause only at dedicated pages (after each task) to get meaningful information about the time on task.

\subsubsection{Scenario}

For each task a fictitious primary school student describes his or her program. To ensure that our results cannot be attributed to differences in the descriptions, the 14 descriptions are structured and filled with content in the same way (an example is shown in \cref{fig:description_example}). In the first two paragraphs, the students says hello and that he or she wants to create an animation or a game. In the following paragraphs the student explains which functions have already been correctly implemented. In the last paragraph the student describes what does not work yet and thus the desired final state. After having repaired and submitted the given program the participants wrote feedback on the bug to the fictitious primary school student. Giving feedback corresponds to the procedure in school and we intended that the teachers in training reflected more on the bug: We attempted that participants repair the program less by trial and error or by following the hint passively but are rather cognitively strongly involved. When participants were not able to repair a program, they still uploaded their final result and instead of feedback on the bug, they were allowed to write feedback on a positive implementation in the student's program instead.

\subsubsection{A/B study}

While course participants were recommended to take part in the programming sessions, it was not mandatory. We therefore assigned participants to \groupA or \groupB on-the-fly through the web interface: Whenever a participant started with Task 1, they were assigned to the smaller of the two groups.
%More precisely, when a participant started with Task 1, we put this participant in the group with less participants and if both groups were the same size, we put the participant in \groupB. This results in splitting up the participants alternately in \groupA and \groupB. 
%as it could be that, e.g., participants with problems rather tend to stop which would lead to apparently better results for later tasks and thus a complicated analysis. 
%This resulted in \numberOfParticipantsCtrl~active participants in \groupA and \numberOfParticipantsTrmt~in \groupB. 
The example task (Task 0) was identical for both groups. The first half of the following tasks (Tasks 1--7) differed between \groupA and \groupB only in that \groupB received one hint for each task. The second half of the tasks (Tasks 8--14) did not differ between the groups: no hints were given for either group. In this way we can examine if there are any effects of using hints within the first tasks and then not providing them within the following tasks that contain the same bug patterns but no hints.

\subsubsection{Bug Patterns, Hints and Tasks}

\input{complexity_table.tex}

For each task, we created or adapted a \scratch program based on projects used previously for teaching. \cref{tab:task_complexity} shows the values of different complexity measures for each program. We modified each of these programs to contain one instance of a bug pattern. 

We decided to use the seven most common bug patterns described by Fraedrich et al.~\cite{fraedrich2020}, who found each of these bug patterns in more than 1,500 \scratch projects of 33,655 publicly shared buggy \scratch projects. By choosing exactly seven bug patterns we made a compromise between a reasonable amount of time and the largest number of bug patterns and thus generalisability. We used the bug patterns in the order shown in \cref{tab:hints}. By this, we attempted to maximise the temporal distance between bug patterns which are related---such as \emph{Message Never Sent} and \emph{Message Never Received} or \emph{Missing Loop Sensing} and \emph{Forever Inside Loop}. 

The hints given to \groupB for Tasks 1 to 7 were generated using \litterbox\footnote{http://scratch-litterbox.org/, last accessed 27.05.2021}. As shown in \cref{fig:hint_example_MCI}, each hint contains both images and text. The text consists of (1) an explanation of the problem and a clarification of the underlying misconception and (2) a generic suggestion on how to remove the bug pattern. The images show the sprite and the script with the highlighted block, where the cause of the bug is located. Table~\ref{tab:hints} also lists the hints for Tasks~1 to 7. As \litterbox analyses the code statically and does not deal with content information about the programming task, some hints provide several options on how to proceed. For the bug pattern \emph{Missing Clone Initialisation} (\cref{fig:hint_example_MCI}), e.g., two different event handlers could solve the bug pattern but only one of them might make sense in the associated program. Thus, the user has to decide which option to use to achieve the intended aim of the program. Consequently, it must be examined to what extent generic hints on bug patterns can support learners.

\subsection{Data Analysis}

We analysed the participants' performance using the criteria (1)~correct fixing of the broken functionalities and (2)~time needed.

\subsubsection{Effectiveness} 
We used a Chi-squared test to measure if the functionality differs significantly between \groupB and \groupA and used odds ratio (\OR) to calculate the effect size. If \OR = 1.0, there are no effects in favour of any group. Values below 1.0 indicate effects in favour of \groupA and values over 1.0 indicate effects in favour of \groupB. % The stronger the value goes in the respective direction, the greater are the effects.

To find out if a broken functionality was correctly fixed we mainly used automated \whisker~\cite{stahlbauer2019testing} tests. \whisker tests \scratch programs dynamically, which means that \whisker runs the programs and checks if the conditions given in the tests are true. We created between \whiskerTestsNumberOfAssertsMin~and \whiskerTestsNumberOfAssertsMax~tests for each task with a median of \whiskerTestsNumberOfAssertsMedian. These tests checked not only if the programs were correctly fixed, but also whether any other functionality of the program was broken. The number of test cases per task differs depending on attributes such as the number of sprites. 

A submitted program was considered correct if (1) the broken functionality was fixed and (2) no other functionality was broken. For three tasks we had to supplement the tests due to limitations of \whisker: As stamps (Task 12) and the lack of delay (Tasks 7 and 14) could not be tracked yet with \whisker at the time of the experiment, we used \litterbox to analyse if the matching solution patterns were implemented and to check whether the bug pattern was removed.

During the creation of the first version of the \whisker tests (and for Tasks 7, 12 and 14 the \litterbox analyses), we checked with individual programs if the automated results are correct. Then, we rated 420 random programs (30 per task) manually and used the \numDeviationsRoundOne~deviations to refine the \whisker tests. After this refinement step we rated another subset of 210 randomly chosen programs (15 per task) manually: This procedure confirms a very good inter-rater reliability ($K$ = \toolEvaluationCompleteCohensKappa) and thus reliable automated results.

\subsubsection{Hint Evaluation}
After submitting their solution for a task participants of \groupB evaluated the hint on a 5-point Likert scale regarding its support with debugging. Participants were then also asked to explain their rating. This provided us with qualitative data, and we analysed the comments using thematic analysis~\cite{bergman2010hermeneutic}: Themes are first collected, then counted and in a final step again related to the original data and our research questions. To ensure inter-rater reliability two raters (one author and one assistant) classified a randomly chosen subcorpus of 35 statements (five per Task 1--7) and agreed on a coding scheme. Then, each rater rated half of the statements and additionally five random statements per task. The comparison of these 35 statements confirms a strong inter-rater agreement with $K$ = \interRaterReliabilityHintEvaluationKappa.

\subsubsection{Efficiency} 
We used a Wilcoxon rank sum test to measure if the time differs significantly between \groupB and \groupA and used the Vargha and Delaney $\hat{A}_{12}$ measure to calculate the effect size. If $ \hat{A}_{12} = .50 $, there are no effects in favour of any group. Values below .50 indicate effects in favour of \groupA and values over .50 indicate effects in favour of \groupB. %The stronger the value goes in the respective direction, the greater are the effects.
The time for completing a task was determined using the website on which we provided the tasks. For each task, we started measuring the time when a participant accessed the page of the respective task (where the task is described and the participants down- and upload the program). Time measurement for the task was stopped when the participants submitted their final program. Time spent for giving feedback to the fictitious student, and for \groupB rating the hint, were not included in the time we tracked.

\subsection{Threats to Validity}

Threats to \emph{external validity} result from our choice of participants, programs and bugs: Our participants consisted only of primary school teachers in training, so effects for teachers at other levels might be different. However, we believe that primary school teachers are one of the most important target groups. We selected a subset of bug patterns and other bug patterns might have other particularities. As we used common bug patterns, many instances of bugs are covered and as we used seven of them, the general insights might be applicable to other bug patterns. Furthermore, our created or adapted programs differ in their complexity. This is why we included the complexity of the programs as a possible explanation of certain results.
Threats to \emph{internal validity} might result from our experiment setup and technical infrastructure: The experiment was embedded in an online course (due to Covid-19 reasons), and participants did not work in a controlled environment. Our time measurement may be unreliable, since other events might interfere (e.g., browser sessions can be accidentally closed).
Threats to \emph{construct validity} arise since we cannot easily measure whether participants have, or can detect, misconceptions. However, we combined several measurements: In this paper, we will evaluate the functionality and time needed. As a next step we will evaluate the feedback the participants gave to the fictitious students.
%
%Furthermore, the static order of the exercises and bug patterns can be seen as an internal threat to validity. Splitting up the groups in more diversified groups regarding exercise and bug pattern order would have made them too small.

%% file: complexity_table.tex
\begin{table}
\centering
\caption{Complexity of the programs.}
\resizebox{\columnwidth}{!}{
\label{tab:task_complexity}
\setlength{\tabcolsep}{0.25em}
\begin{tabular}{rrrrrrrrrrr}
\toprule
 Task &  Blocks &  Scripts &  Sprites &   HD$^{1}$ &      HE$^{2}$ &  HL$^{3}$ &  HS$^{4}$ &  HV$^{5}$ &  ICC$^{6}$ &  WMC$^{7}$ \\
\midrule
    1 &        81 &         10 &          3 & 27.4 & 19485.3 & 118 &  65 & 710 &   24 &   22 \\
    2 &        44 &          6 &          2 & 20.0 &  7657.1 &  71 &  42 & 382 &   13 &   11 \\
    3 &        34 &          5 &          2 & 14.7 &  4270.7 &  54 &  42 & 291 &   10 &    9 \\
    4 &        52 &          6 &          2 & 28.0 & 12609.4 &  79 &  52 & 450 &   15 &   14 \\
    5 &        47 &          7 &          4 & 14.9 &  6326.6 &  74 &  53 & 423 &   13 &   12 \\
    6 &        36 &          5 &          4 & 11.6 &  3669.6 &  60 &  39 & 317 &   12 &   10 \\
    7 &        62 &          9 &          4 & 20.2 & 12275.4 & 102 &  62 & 607 &   17 &   16 \\
    8 &        56 &         10 &          5 & 16.2 &  9808.4 & 107 &  50 & 603 &   14 &   15 \\
    9 &        49 &          6 &          4 & 15.2 &  5709.8 &  70 &  41 & 375 &   13 &   14 \\
   10 &        66 &          8 &          6 & 18.9 & 12093.2 & 109 &  58 & 638 &   19 &   19 \\
   11 &        77 &         13 &          4 & 18.3 & 13536.6 & 125 &  60 & 738 &   26 &   23 \\
   12 &        61 &          7 &          2 & 25.4 & 13302.1 &  95 &  46 & 524 &   15 &   17 \\
   13 &        38 &          3 &          2 & 15.8 &  4926.9 &  63 &  31 & 312 &    8 &    8 \\
   14 &        39 &          7 &          2 & 16.0 &  5543.9 &  65 &  40 & 345 &   13 &   11 \\
   \bottomrule
   \multicolumn{11}{l}{$^{1}$Halstead Difficulty \cite{halstead1977elements} $^{2}$Halstead Effort \cite{halstead1977elements} $^{3}$Halstead Length \cite{halstead1977elements}} \\
   
   \multicolumn{11}{l}{$^{4}$Halstead Size \cite{halstead1977elements} $^{5}$Halstead Volume \cite{halstead1977elements}} \\
   
   \multicolumn{11}{l}{$^{6}$Interprocedural Cyclomatic Complexity $^{7}$Weighted Method Count} \\
 
\end{tabular}
}
\end{table}

%% file: content/results.tex
\section{Results}
\label{sec:resuls}

\subsection{RQ1: Effects of Showing Hints}
To answer RQ1 we consider Tasks 1 to 7, where participants of \groupA had to work without hints and participants of \groupB were shown hints. 
\Cref{fig:hint_evaluation_quantitative} and \cref{tab:hint_evaluation} deal with the hint evaluation: \Cref{fig:hint_evaluation_quantitative} shows how the participants rated the hint on a five-point Likert scale. Participants were asked to explain their rating and \cref{tab:hint_evaluation} shows how often different aspects of the hints were mentioned.

\begin{figure}[tb]
	\centering
	\includegraphics[width=\columnwidth]{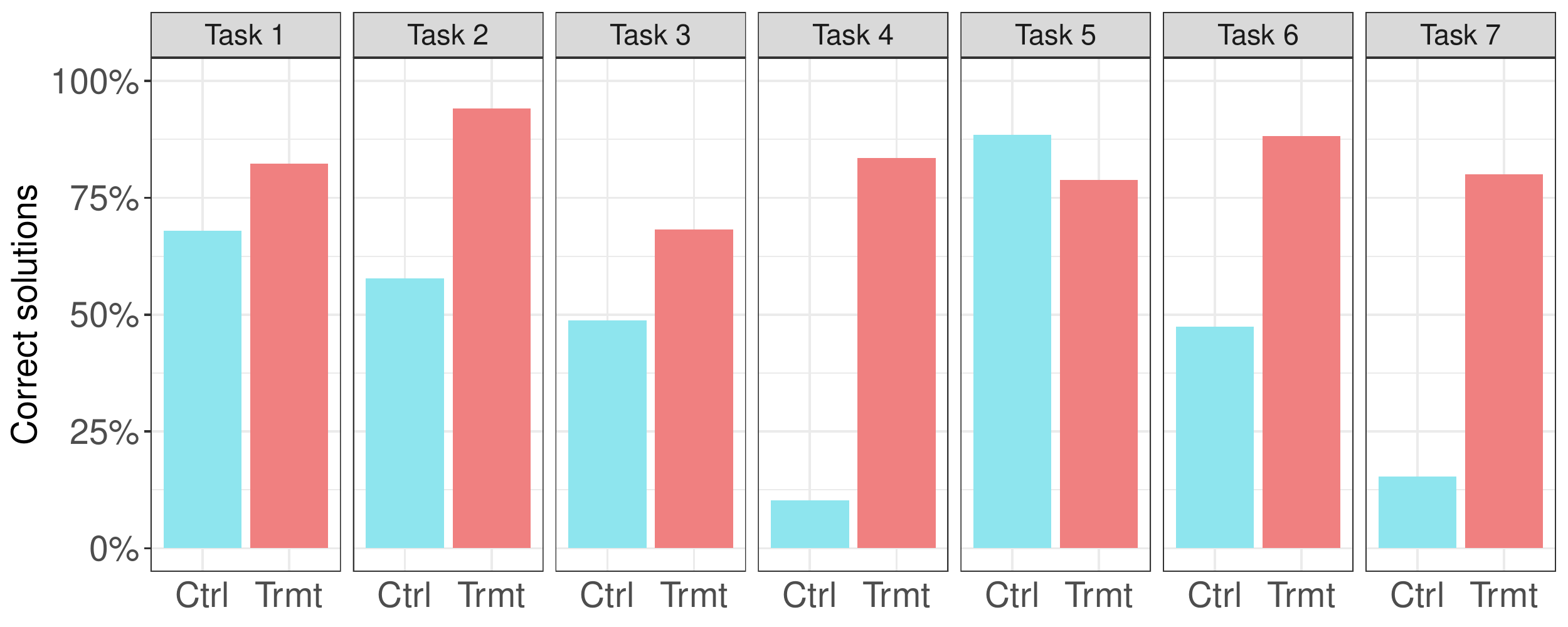} 
	\vspace{-2.2em}
	\caption{\label{fig:points_1-7}Proportion of correct solutions for Tasks 1 to 7.}
%	\vspace{-0.5em}
\end{figure}

\begin{table}[t]\centering
	\caption{Effect sizes for the functionality of Tasks 1 to 7. \\ \textmd{Effect sizes associated with significant $p$-values ($p<.05$) are bold.}}
	\label{tab:p-values_1-7}
	\vspace{-0.8em}
	\setlength\tabcolsep{0.1cm}
	\resizebox{\columnwidth}{!}{
\begin{tabular}{rrrrrrrr} 
	\toprule
	& Task 1 & Task 2 & Task 3 & Task 4 & Task 5 & Task 6 & Task 7 \\
	& \OR & \OR & \OR & \OR & \OR & \OR & \OR \\
	\midrule
All & $ \taskIEffectSize $ &  $ \textbf{\taskIIEffectSize} $ & $ \textbf{\taskIIIEffectSize} $ & $ \textbf{\taskIVEffectSize} $ & $ \taskVEffectSize $ & $ \textbf{\taskVIEffectSize} $ & $ \textbf{\taskVIIEffectSize} $ \\
Experienced & $ \taskIExperienceEffectSize $ & $ \textbf{\taskIIExperienceEffectSize} $  & $ \taskIIIExperienceEffectSize $ &  $ \textbf{\taskIVExperienceEffectSize} $ &  $ \textbf{\taskVExperienceEffectSize} $ & $ \textbf{\taskVIExperienceEffectSize} $  & $ \textbf{\taskVIIExperienceEffectSize} $ \\
Inexperienced &  $ \taskINoExperienceEffectSize $ & $ \textbf{\taskIINoExperienceEffectSize} $ &  $ \textbf{\taskIIINoExperienceEffectSize} $ &  $ \textbf{\taskIVNoExperienceEffectSize} $ &  $ \taskVNoExperienceEffectSize $ &  $ \textbf{\taskVINoExperienceEffectSize} $  & $ \textbf{\taskVIINoExperienceEffectSize} $ \\
\bottomrule
\end{tabular}
}
\end{table}

\subsubsection{Effectiveness: Differences in Terms of Functionality}
\Cref{fig:points_1-7} shows how many correct solutions were produced per task and per group and \Cref{tab:p-values_1-7} shows the effect sizes for each task for all, experienced and inexperienced participants. 
%\paragraph{General Differences}
Overall, \groupB submitted \percentageFunctionalityDifferenceLower\% more correct solutions than \groupA (\groupB: \percentageFunctionalityTrmtLower\%; \groupA: \percentageFunctionalityCtrlLower\%). With an odds ratio of \tasksLowerEffectSizeMean~on average over all Tasks 1 to 7 there are medium effects in favour of the hints. 
%
%Considering participants as effective in debugging if they repaired at least five of the seven programs, also shows the clear difference between \groupB and \groupA: \percentageFiveOrMoreCorrectTasksTrmtLower\%
%of \groupB submitted five or more correct programs in contrary to \percentageFiveOrMoreCorrectTasksCtrlLower\%
%of \groupA. Thus, using hints leads to more than three times as many effective debuggers as debugging without hints. 

%Nevertheless, not all participants of \groupB produced correct solutions. This matches the hint evaluation, where some comments contained negative statements (\cref{tab:hint_evaluation}). \textbf{This shows that there are factors hindering the effects of hints.} 

%\subsubsection{Differences between the Tasks in Terms of Functionality}

%\paragraph{Differences between the Tasks}
We note substantial variation between the individual tasks (\cref{fig:points_1-7}): The hints on Tasks 2, 3, 4, 6 and 7 lead to \groupB producing significantly more correct solutions ($\taskIIChiSquarePValue$) with Task 7 having the highest effect size ($\OR = \taskVIIEffectSize$). 
The differences for Tasks 1 and 5 however are not significant. For Task 1, it is barely not significant in favour of the hints ($\taskIChiSquarePValue$, $\OR = \taskIEffectSize$). For Task 5, the percentage of correct solutions is even lower for \groupB (\percentageCorrectSolutionsTrmtTaskV\%) than for \groupA (\percentageCorrectSolutionsCtrlTaskV\%) though not significantly ($\taskVChiSquarePValue$, $\OR = \taskVEffectSize$).

%Although the hint leads to a significant improvement for Task 3, the odds ratio is lower compared to the other significant tasks ($\taskIIIChiSquarePValue$, $\OR = \taskIIIEffectSize$). Indeed, with \percentageCorrectSolutionsTrmtTaskIII\% for no other task of the Tasks 1 to 7, the percentage of correct programs by \groupB is lower (\cref{fig:points_1-7}). 
%

%\subsubsection{Differences between Experienced and Inexperienced Participants in Terms of Functionality}

%\paragraph{Differences between (In-)Experienced Participants}
\Cref{tab:p-values_1-7} suggests that showing hints generally leads to more correct programs for both experienced and inexperienced participants, but the effects are larger for experienced participants (experienced: $\OR = \tasksLowerExperienceEffectSizeMean$; inexperienced: $\OR = \tasksLowerNoExperienceEffectSizeMean$). 
%
%Before teaching a topic at school, teachers prepare themselves: If they did not acquire competencies in their studies or an additional advanced training, they will usually acquire skills through self-study. Consequently, teachers in practice have at least some programming experience when they teach programming. \textbf{Our results indicate that especially experienced participants and thus teachers in practice should use tool-support to recognise bug patterns.}
%
%However, both experienced and inexperienced participants benefit from hints but
Only for the Tasks 3 and 5, experienced participants benefit less than inexperienced participants. For Task 3, only inexperienced participants of \groupB submitted significantly more correct programs than those of \groupA ($\taskIIINoExperienceChiSquarePValue$, $\OR = \taskIIINoExperienceEffectSize$).
%, while the experienced participants of \groupB did not submit significantly more correct tasks than those of \groupA ($\taskIIIExperienceChiSquarePValue$, $\OR = \taskIIIExperienceEffectSize$).
%
For Task 5---the only task with slightly negative effects of hints when considering all participants---, inexperienced participants of \groupB did \emph{not} produce significantly more or less correct programs than those of \groupA ($\taskVNoExperienceChiSquarePValue$, $\OR = \taskVNoExperienceEffectSize$), but the experienced participants of \groupB produced significantly fewer correct solutions than those of \groupA ($\taskVExperienceChiSquarePValue$, $\OR = \taskVExperienceEffectSize$).

% Efficiency
\subsubsection{Efficiency: Differences in Terms of Time}

\begin{figure}[tb]
	\centering
	\includegraphics[width=\columnwidth]{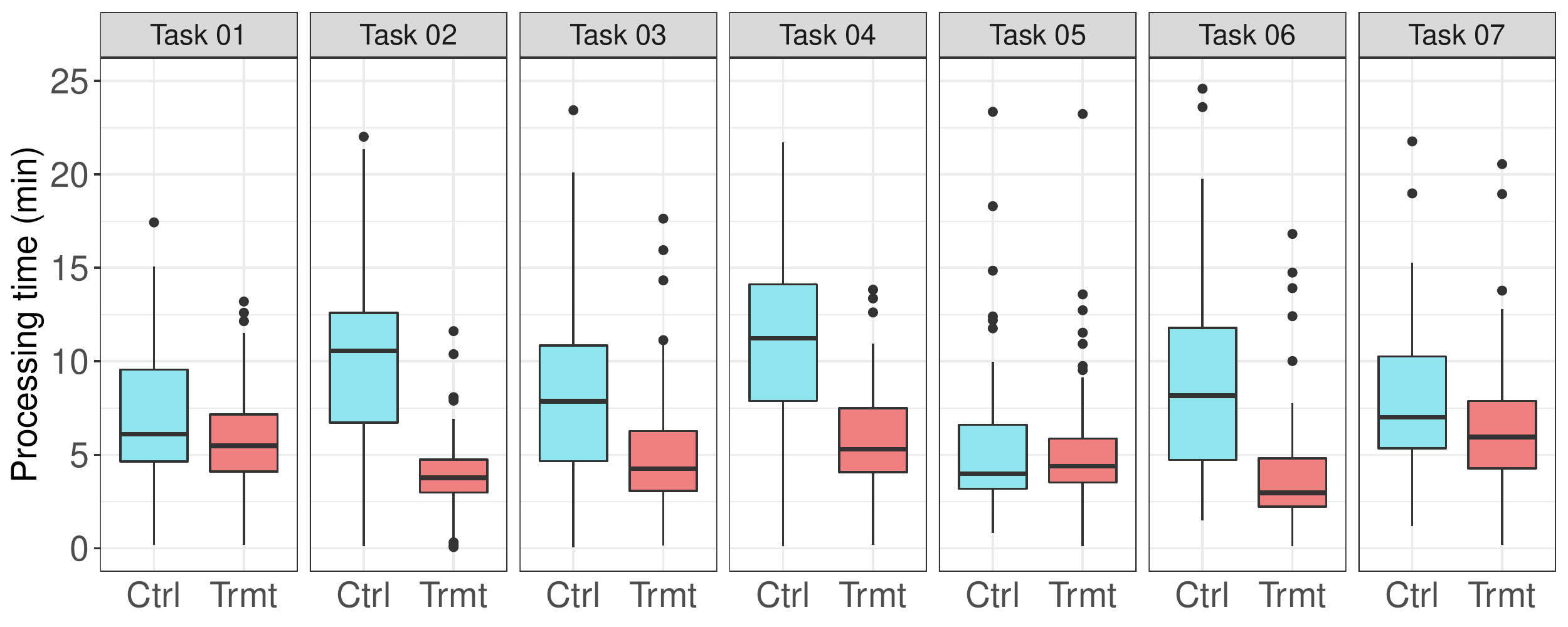} 
	\vspace{-2.2em}
	\caption{\label{fig:time_1-7}Time needed to submit a program for Tasks 1 to 7.}
%	\vspace{-0.5em}
\end{figure}

\begin{table}[t]\centering
	\caption{Effect sizes for the time needed for Tasks 1 to 7. \\ \textmd{Effect sizes associated with significant $p$-values ($p<.05$) are bold.}}
	\label{tab:p-values_time_1-7}
	\vspace{-0.8em}
	\setlength\tabcolsep{0.07cm}
	\resizebox{\columnwidth}{!}{
\begin{tabular}{rrrrrrrr}
	\toprule
	& Task 1 & Task 2 & Task 3 & Task 4 & Task 5 & Task 6 & Task 7 \\
	& $ \hat{A}_{12} $ & $ \hat{A}_{12} $ & $ \hat{A}_{12} $ & $ \hat{A}_{12} $ & $ \hat{A}_{12} $ & $ \hat{A}_{12} $ & $ \hat{A}_{12} $  \\
\midrule
All & $\textbf{\processingTimeTaskIEffectSize}$ & $\textbf{\processingTimeTaskIIEffectSize}$ & $\textbf{\processingTimeTaskIIIEffectSize}$ & $\textbf{\processingTimeTaskIVEffectSize}$ & $\processingTimeTaskVEffectSize$ & $\textbf{\processingTimeTaskVIEffectSize}$ & $\textbf{\processingTimeTaskVIIEffectSize}$\\
%
% shortened: Correct & \delG{\processingTimeJustCorrectTaskIWilcoxPValue} & $\processingTimeJustCorrectTaskIEffectSize$ & \delP{\processingTimeJustCorrectTaskIIWilcoxPValue}{}^* & $\processingTimeJustCorrectTaskIIEffectSize$ & \delP{\processingTimeJustCorrectTaskIIIWilcoxPValue}{}^* & $\processingTimeJustCorrectTaskIIIEffectSize$ & \delG{\processingTimeJustCorrectTaskIVWilcoxPValue} & $\processingTimeJustCorrectTaskIVEffectSize$ & \delG{\processingTimeJustCorrectTaskVWilcoxPValue} & $\processingTimeJustCorrectTaskVEffectSize$ & \delP{\processingTimeJustCorrectTaskVIWilcoxPValue}{}^* & $\processingTimeJustCorrectTaskVIEffectSize$ & \delG{\processingTimeJustCorrectTaskVIIWilcoxPValue} & $\processingTimeJustCorrectTaskVIIEffectSize$\\
%
% shortened: Incorrect & \delG{\processingTimeJustWrongTaskIWilcoxPValue}{}^* & $\processingTimeJustWrongTaskIEffectSize$ & \delG{\processingTimeJustWrongTaskIIWilcoxPValue}{}^* & $\processingTimeJustWrongTaskIIEffectSize$ & \delG{\processingTimeJustWrongTaskIIIWilcoxPValue}{}^* & $\processingTimeJustWrongTaskIIIEffectSize$ & \delG{\processingTimeJustWrongTaskIVWilcoxPValue}{}^* & $\processingTimeJustWrongTaskIVEffectSize$ & \delG{\processingTimeJustWrongTaskVWilcoxPValue} & $\processingTimeJustWrongTaskVEffectSize$ & \delG{\processingTimeJustWrongTaskVIWilcoxPValue}{}^* & $\processingTimeJustWrongTaskVIEffectSize$ & \delG{\processingTimeJustWrongTaskVIIWilcoxPValue} & $\processingTimeJustWrongTaskVIIEffectSize$\\
%
Experienced & $\processingTimeExperienceTaskIEffectSize$ &  $\textbf{\processingTimeExperienceTaskIIEffectSize}$ &  $\textbf{\processingTimeExperienceTaskIIIEffectSize}$ &  $\textbf{\processingTimeExperienceTaskIVEffectSize}$ &  $\processingTimeExperienceTaskVEffectSize$ &  $\textbf{\processingTimeExperienceTaskVIEffectSize}$ &  $\textbf{\processingTimeExperienceTaskVIIEffectSize}$ \\
Inexperienced & $\processingTimeNoExperienceTaskIEffectSize$ &  $\textbf{\processingTimeNoExperienceTaskIIEffectSize}$ &  $\textbf{\processingTimeNoExperienceTaskIIIEffectSize}$ &  $\textbf{\processingTimeNoExperienceTaskIVEffectSize}$ &  $\processingTimeNoExperienceTaskVEffectSize$ &  $\textbf{\processingTimeNoExperienceTaskVIEffectSize}$ &  $\processingTimeNoExperienceTaskVIIEffectSize$ \\
\bottomrule
\end{tabular}
}
\end{table}

\Cref{fig:time_1-7} shows the time required by the participants per task and per group and \cref{tab:p-values_time_1-7} shows the respective effect sizes for all, experienced and inexperienced participants.
%
%\paragraph{General Differences}
On average \groupB needs \meanProcessingTimeTrmtLower~minutes to submit a program and \groupA needs \meanProcessingTimeCtrlLower~minutes. Thus, the effort of debugging is reduced by \meanProcessingTimeDifferenceLower~minutes. The effect size of $\hat{A}_{12}$ = \processingTimeLowerEffectSizeMean~on average over all Tasks 1 to 7 indicates medium effects in favour of the hints.

Looking at individual tasks, results are similar to those regarding functionality: The hint on Task 5 again slightly impairs the performance, but again not significantly (\processingTimeTaskVWilcoxPValue, $\hat{A}_{12}$ = \processingTimeTaskVEffectSize). 
A slight difference to the results of the functionality is that the hint on Task 1
%---which was barely not significant for the functionality---
is significant in terms of time (\processingTimeTaskIWilcoxPValue, $\hat{A}_{12}$ = \processingTimeTaskIEffectSize). %This can be explained by the incorrect programs (\processingTimeJustWrongTaskIWilcoxPValue, $\hat{A}_{12}$ = \processingTimeJustWrongTaskIEffectSize). 

%Summarized, the time needed depends mostly on the functionality results of the respective task. \textbf{This indicates that showing
%hints has to be effective to be efficient.}

%\subsubsection{Differences Between Experienced and Inexperienced Participants in Terms of Time}
%\paragraph{Differences between (In-)Experienced Participants}
The effect sizes in terms of time are very similar for inexperienced and experienced participants (experienced: $\hat{A}_{12}$ = \processingTimeLowerExperienceEffectSizeMean; inexperienced: $\hat{A}_{12}$ = \processingTimeLowerNoExperienceEffectSizeMean). %Indeed, being (in-)experienced only slightly influences the time needed both for \groupB (experienced: \meanProcessingTimeExperienceTrmtLower~minutes; inexperienced: \meanProcessingTimeNoExperienceTrmtLower~minutes) and \groupA (experienced: \meanProcessingTimeExperienceCtrlLower~minutes; inexperienced: \meanProcessingTimeNoExperienceCtrlLower~minutes). 
%Our results suggest that receiving a hint hardly changes that the prior programming experience does not have a considerable impact on the time needed. \textbf{In general, hints help inexperienced and experienced participants equally in terms of time.}
%
However, for Task 7, only the experienced participants of \groupB submitted their programs significantly earlier than those of \groupA (\processingTimeExperienceTaskVIIWilcoxPValue, $\hat{A}_{12}$ = \processingTimeExperienceTaskVIIEffectSize).%, whereas there is no significant difference between \groupB and \groupA for inexperienced participants (\processingTimeNoExperienceTaskVIIWilcoxPValue, $\hat{A}_{12}$ = \processingTimeNoExperienceTaskVIIEffectSize). % For Task 1, only experienced participants submitted almost significantly more correct programs (\processingTimeExperienceTaskIWilcoxPValue, $\hat{A}_{12}$ = \processingTimeExperienceTaskIEffectSize).   \\

\summary{RQ1}{Hints on bug patterns improve the performance at debugging and fixing programs: In general, programs can be fixed significantly more often and faster. In our experiment, one hint 
%on the bug pattern  %\emph{Message Never Received}---where the origin of the bug is shown but not the point where the bug itself is---
significantly impeded the debugging performance regarding the effectiveness of experienced participants.} 
\newline

%-------------------------------------------------------------------------------
\subsection{RQ2: Effects of Knowing Patterns %Effects of Transferring Hints
}
To answer RQ2 we consider Tasks 8 to 14, where participants of both groups had to work without hints.

\subsubsection{Effectiveness: Differences in Terms of Functionality}

\begin{figure}[tb]
	\centering
	\includegraphics[width=\columnwidth]{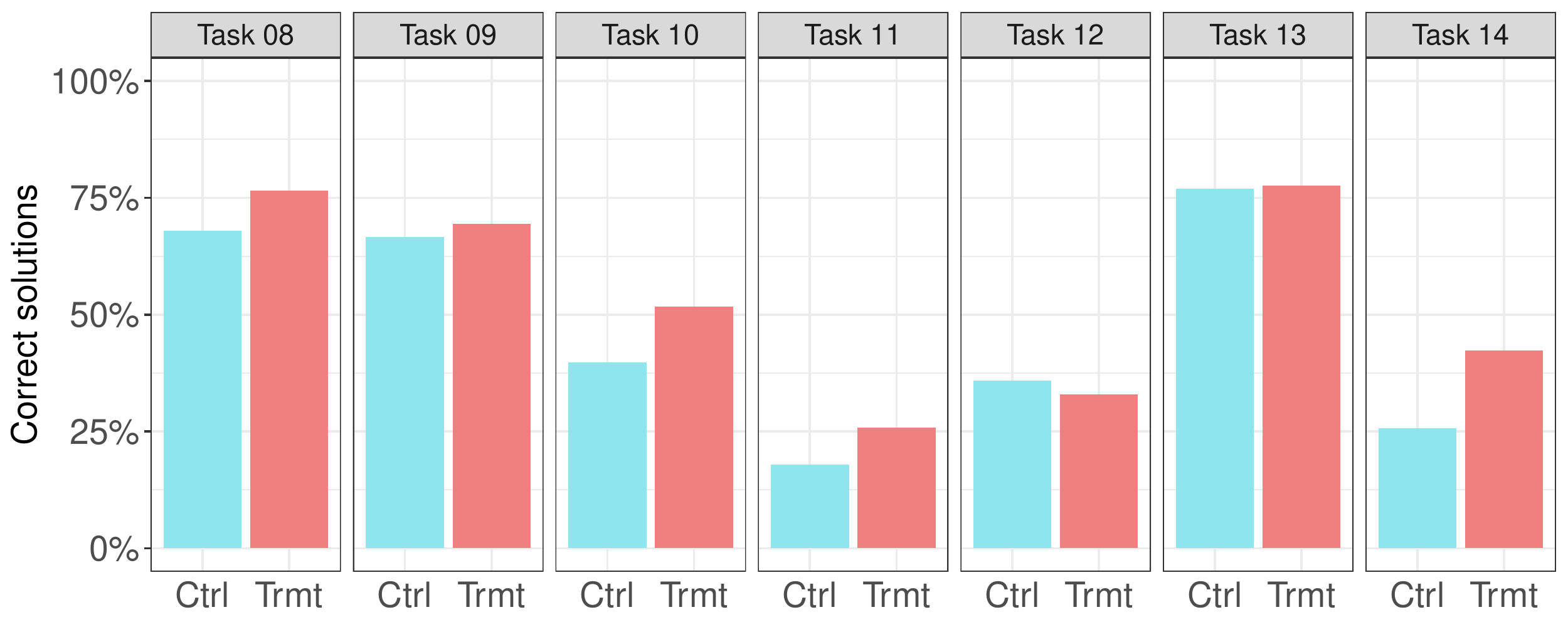} 
	\vspace{-2.2em}
	\caption{\label{fig:points_8-14}Proportion of correct solutions for Tasks 8 to 14.}
%	\vspace{-0.5em}
\end{figure}

\begin{table}[t]\centering
	\caption{Effect sizes for the functionality of Tasks 8 to 14. \\ \textmd{Effect sizes associated with significant $p$-values ($p<.05$) are bold.}}
	\label{tab:p-values_8-14}
	\vspace{-0.8em}
	\setlength\tabcolsep{0.07cm}
	\resizebox{\columnwidth}{!}{
	\begin{tabular}{rrrrrrrr} 
		\toprule
		& Task 8 & Task 9 & Task 10 & Task 11 & Task 12 & Task 13 & Task 14 \\
		& \OR & \OR & \OR & \OR & \OR & \OR & \OR \\
		\midrule
		All & $ \taskVIIIEffectSize $ &  $ \taskIXEffectSize $ &  $ \taskXEffectSize $ &  $ \taskXIEffectSize $ &  $ \taskXIIEffectSize $ &  $ \taskXIIIEffectSize $ &   $ \textbf{\taskXIVEffectSize} $ \\
		Experienced & $ \taskVIIIExperienceEffectSize $ & $ \taskIXExperienceEffectSize $ & $ \taskXExperienceEffectSize $ & $ \taskXIExperienceEffectSize $ & $ \taskXIIExperienceEffectSize $ &  $ \taskXIIIExperienceEffectSize $ &  $ \taskXIVExperienceEffectSize $ \\
		Inexperienced & $ \taskVIIINoExperienceEffectSize $ &  $ \taskIXNoExperienceEffectSize $ &  $ \taskXNoExperienceEffectSize $ &  $ \taskXINoExperienceEffectSize $ &  $ \taskXIINoExperienceEffectSize $ &  $ \taskXIIINoExperienceEffectSize $ &  $ \taskXIVNoExperienceEffectSize $ \\
		\bottomrule
	\end{tabular}
	}
\end{table}

\Cref{fig:points_8-14} shows how many correct solutions were handed in per task and per group. \Cref{tab:p-values_8-14} shows the effect sizes for each task and it differentiates between experienced and inexperienced participants. 
%
%\paragraph{General Differences}
Comparing both groups, we observe that \groupB has only a slight edge over \groupA:
%The differences between \groupA and \groupB are far smaller than for the Tasks 1 to 7, although we nevertheless observe a slight improvement:
%
\groupB submitted \percentageFunctionalityTrmtHigher\% and \groupA \percentageFunctionalityCtrlHigher\% correct programs in total over Task 8 to 14 and we note an average effect size of \tasksHigherEffectSizeMean.
%Thus, the effectiveness is only increased slightly 
%by \percentageFunctionalityDifferenceLower\% 
%Considering participants as effective in debugging if they repaired at least five of the seven programs, also shows slight differences between \groupB and \groupA: \percentageFiveOrMoreCorrectTasksTrmtHigher\% of \groupB submitted five or more correct programs in contrary to \percentageFiveOrMoreCorrectTasksCtrlHigher\% of \groupA.
%Despite the small effect size of \tasksHigherEffectSizeMean~on average over all Tasks 8 to 14, it still is comforting, that 
%
% shortened: Nevertheless, this can be seen in a more favourable light when taking into account that direct instruction compared to discovery-based learning may even lead to worse performance when completing transfer tasks \cite{kapur2012designing}.

%This can be seen as a slightly comforting result as providing information may also lead to worse performance than discovery learning when completing transfer tasks \cite{kapur2012designing}. %We conjecture that this result can be attributed to the hints containing not only a suggestion on how to remove the bug pattern, but also an explanation of the problem and a clarification of the underlying misconception. 

%\paragraph{Differences between the Tasks}
Looking at the individual tasks (\cref{fig:points_8-14}), Task 14 clearly stands out, because \groupB produced significantly more correct solutions for this task only
%only Task 14 shows a significant difference, with \groupB producing more correct solutions 
($\taskXIVChiSquarePValue$, $\OR = \taskXIVEffectSize$). 
%
%It appears that the corresponding hint on the bug pattern \emph{Stuttering Movement} from Task 7 can quite easily be transferred to similar problems.  
A second task that stands out in \cref{fig:points_8-14} is Task~12 where \groupA performs even slightly better than \groupB ($\taskXIIChiSquarePValue$, $\OR = \taskXIIEffectSize$). 
%appears to have had a small negative effect on participants facing the same bug pattern in Task 12 without hint, although not significantly so ($\taskXIIChiSquarePValue$, $\OR = \taskXIIEffectSize$).
%
%For the remaining tasks, there is no significant difference. %\groupB generally submitted slightly more correct solutions, although the difference is not significant for any of the tasks. 

% shortened: The effect is largest for Task 10 ($\taskXChiSquarePValue$, $\OR = \taskXEffectSize$), which may be an effect of participants achieving some understanding of the underlying \emph{Comparing Literals} bug from the corresponding hint on Task 3: \enquote{Though, the hint is written unclear and difficult to understand, by reading it several times the hint was very helpful.} (P20). Maybe participants had to think quite a bit about the information provided in the hint for Task 3 %, such that they processed it more deeply and therefore also could transfer it slightly better than \groupA.
%
% shortened: On the other hand, the large improvement seen for the \emph{Forever Inside Loop} bug in Task 6 was not transferred to Task 13 ($\taskXIIIChiSquarePValue$, $\OR = \taskXIIIEffectSize$), which may be a result of the underlying program being rather easy as suggested by different complexity measures (\cref{tab:task_complexity}). Processing patterns sufficiently might be less essential when dealing with simple programs. 
%---which is slightly improved by the hints for other tasks---
%
%\paragraph{Differences between (In-)Experienced Participants}
Inexperienced participants of \groupB show a slightly larger improvement on average when given hints during training than experienced ones (experienced: $\OR = \tasksHigherExperienceEffectSizeMean$; inexperienced: $\OR = \tasksHigherNoExperienceEffectSizeMean$).
%
%This shows that the inexperienced participants achieved better learning using the hints than on their own without hints.
%
However, for Tasks 10 and 11, the effects are noteworthy for experienced participants and somewhat invisible for inexperienced participants (\cref{tab:p-values_8-14}). %Indeed, the effects of hints on experienced participants are almost significant for Task 10 ($\taskXExperienceChiSquarePValue$, $\OR = \taskXExperienceEffectSize$). %Note that both \groupA and \groupB produced the fewest correct solutions for Task 11.    

%which may lead to inexperienced learners of both \groupA and \groupB being unable to repair the bug, especially for Task 11 (\cref{fig:points_8-14}). 
%
%Even though the inexperienced participants of \groupB might have understood the hint, they did not get to the point where they could apply their knowledge. Experienced participants, however, seem to keep track of the multiple sprites and blocks and thus can benefit from having received a hint in the corresponding Tasks 3 and 4. 
%To sum up, our results indicate that the effects of knowing patterns might depend on prior programming experience and might be influenced by the complexity of the program, quality of the hint and inherent bug pattern instance variability.

\subsubsection{Efficiency: Differences in Terms of Time}

\begin{figure}[tb]
	\centering
	\includegraphics[width=\columnwidth]{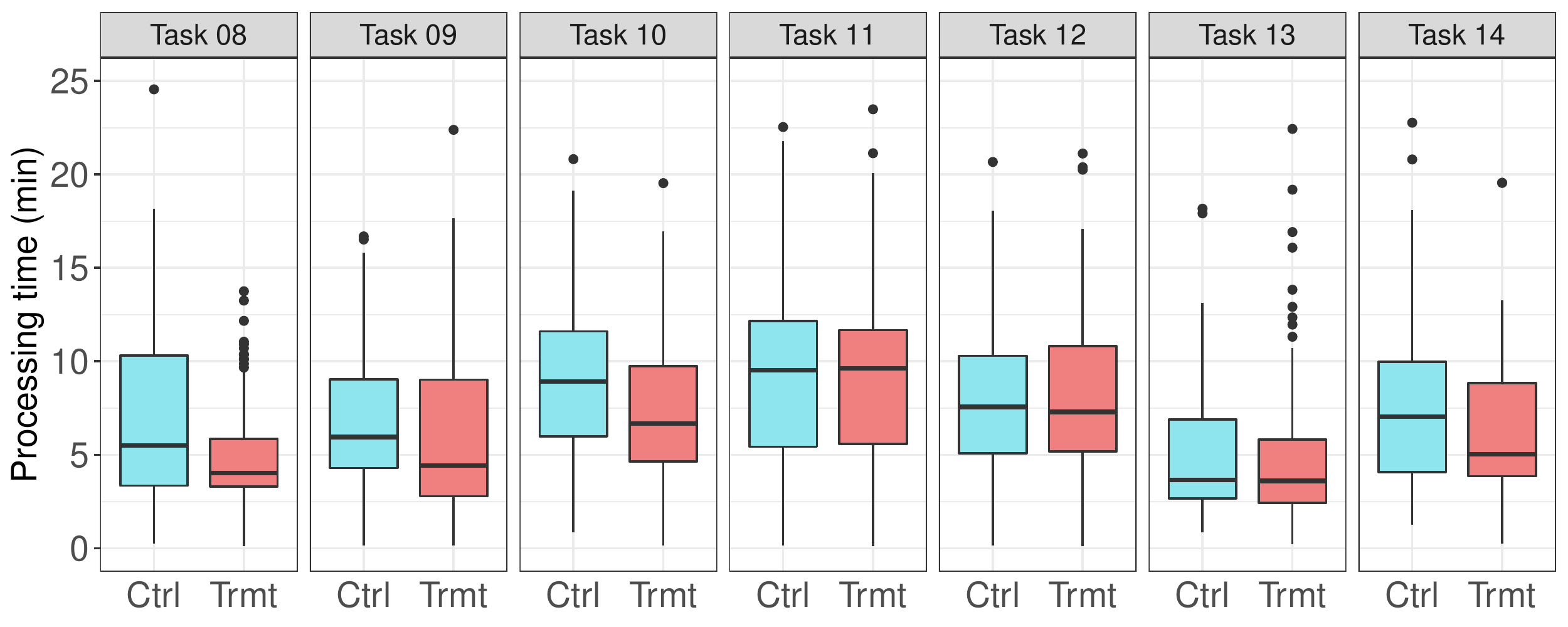} 
	\vspace{-2.2em}
	\caption{\label{fig:time_8-14}Time needed to submit a program for Tasks 8 to 14.}
%	\vspace{-0.5em}
\end{figure}

\begin{table}[t]\centering
	\caption{Effect sizes for the time needed for Tasks 8 to 14. \\ \textmd{Effect sizes associated with significant $p$-values ($p<.05$) are bold.}}
	\label{tab:p-values_time_8-14}
	\vspace{-0.8em}
	\setlength\tabcolsep{0.07cm}
	\resizebox{\columnwidth}{!}{
	\begin{tabular}{rrrrrrrr} 
		\toprule
		& Task 8 & Task 9 & Task 10 & Task 11 & Task 12 & Task 13 & Task 14 \\
		& $ \hat{A}_{12} $ & $ \hat{A}_{12} $ & $ \hat{A}_{12} $ & $ \hat{A}_{12} $ & $ \hat{A}_{12} $ & $ \hat{A}_{12} $ & $ \hat{A}_{12} $ \\
		\midrule
		All & $\textbf{\processingTimeTaskVIIIEffectSize}$ &  $\processingTimeTaskIXEffectSize$ &  $\textbf{\processingTimeTaskXEffectSize}$ &  $\processingTimeTaskXIEffectSize$ &  $\processingTimeTaskXIIEffectSize$ &  $\processingTimeTaskXIIIEffectSize$ &  $\processingTimeTaskXIVEffectSize$\\
		%
		% shortened: Correct & \delG{\processingTimeJustCorrectTaskVIIIWilcoxPValue} & $\processingTimeJustCorrectTaskVIIIEffectSize$ & \delG{\processingTimeJustCorrectTaskIXWilcoxPValue}{}^* & $\processingTimeJustCorrectTaskIXEffectSize$ & \delG{\processingTimeJustCorrectTaskXWilcoxPValue} & $\processingTimeJustCorrectTaskXEffectSize$ & \delG{\processingTimeJustCorrectTaskXIWilcoxPValue} & $\processingTimeJustCorrectTaskXIEffectSize$ & \delG{\processingTimeJustCorrectTaskXIIWilcoxPValue} & $\processingTimeJustCorrectTaskXIIEffectSize$ & \delG{\processingTimeJustCorrectTaskXIIIWilcoxPValue} & $\processingTimeJustCorrectTaskXIIIEffectSize$ & \delG{\processingTimeJustCorrectTaskXIVWilcoxPValue} & $\processingTimeJustCorrectTaskXIVEffectSize$\\
		%
		% shortened: Incorrect & \delG{\processingTimeJustWrongTaskVIIIWilcoxPValue}{}^* & $\processingTimeJustWrongTaskVIIIEffectSize$ & \delG{\processingTimeJustWrongTaskIXWilcoxPValue} & $\processingTimeJustWrongTaskIXEffectSize$ & \delG{\processingTimeJustWrongTaskXWilcoxPValue} & $\processingTimeJustWrongTaskXEffectSize$ & \delG{\processingTimeJustWrongTaskXIWilcoxPValue} & $\processingTimeJustWrongTaskXIEffectSize$ & \delG{\processingTimeJustWrongTaskXIIWilcoxPValue} & $\processingTimeJustWrongTaskXIIEffectSize$ & \delG{\processingTimeJustWrongTaskXIIIWilcoxPValue} & $\processingTimeJustWrongTaskXIIIEffectSize$ & \delG{\processingTimeJustWrongTaskXIVWilcoxPValue} & $\processingTimeJustWrongTaskXIVEffectSize$\\
        %
        Experienced & $\processingTimeExperienceTaskVIIIEffectSize$ &  $\processingTimeExperienceTaskIXEffectSize$ &  $\textbf{\processingTimeExperienceTaskXEffectSize}$ &  $\processingTimeExperienceTaskXIEffectSize$ &  $\processingTimeExperienceTaskXIIEffectSize$ &  $\processingTimeExperienceTaskXIIIEffectSize$ &  $\processingTimeExperienceTaskXIVEffectSize$ \\
        Inexperienced & $\textbf{\processingTimeNoExperienceTaskVIIIEffectSize}$ &  $\textbf{\processingTimeNoExperienceTaskIXEffectSize}$ &   $\processingTimeNoExperienceTaskXEffectSize$ &  $\processingTimeNoExperienceTaskXIEffectSize$ &  $\processingTimeNoExperienceTaskXIIEffectSize$ &  $\processingTimeNoExperienceTaskXIIIEffectSize$ & $\processingTimeNoExperienceTaskXIVEffectSize$ \\
    \bottomrule
	\end{tabular}}
\end{table}

\Cref{fig:time_8-14} shows the time required by participants per task and per group and \cref{tab:p-values_time_8-14} shows the effect sizes for all, experienced and inexperienced participants.
%
%\paragraph{General Differences}
Overall, it took participants of \groupB \meanProcessingTimeTrmtHigher~minutes to submit a program, while \groupA needed \meanProcessingTimeCtrlHigher~minutes. Thus, the effort of debugging is slightly reduced 
%by \meanProcessingTimeDifferenceLower~minutes 
with an effect size of \processingTimeHigherEffectSizeMean~on average over Tasks 8 to 14.

%Didn't really get the point of this:
%This suggests that transferring learning chances with hints leads to an only slightly improved efficiency with an effect size of $\hat{A}_{12}$ = \processingTimeHigherEffectSizeMean~on average over all Tasks 8 to 14. This may result from debugging independently: When transferring the hint, \groupB also has to locate the bug which takes some time and was not necessary when hints were shown for Tasks 1 to 7. 
%
%
% There are also only small effects for both correct ($\hat{A}_{12}$ = \processingTimeHigherJustCorrectEffectSizeMean for) and incorrect ($\hat{A}_{12}$  = \processingTimeHigherJustWrongEffectSizeMean) programs. Using correct programs only, \groupB needs \meanProcessingTimeJustCorrectTrmtHigher~minutes to submit a program and \groupA needs \meanProcessingTimeJustCorrectCtrlHigher~minutes. When considering only time data of incorrect programs both \groupB and \groupA need about \meanProcessingTimeJustWrongTrmtHigher~minutes to submit a program.  

%\subsubsection{Differences between the Tasks in Terms of Time}
%
%\paragraph{Differences between the Tasks}
The largest effects can be seen in Tasks 8, 9, 10 and 14 although the difference is only significant for Tasks 8 (\processingTimeTaskVIIIWilcoxPValue, $\hat{A}_{12}$ = \processingTimeTaskVIIIEffectSize) and 10 (\processingTimeTaskXWilcoxPValue, $\hat{A}_{12}$ = \processingTimeTaskXEffectSize). % shortened: Interestingly, for Task 8 the time improvements are significantly larger for those participants who submitted an incorrect solution, while for Task 9 the time improvements are significantly larger for participants who submitted a correct solution. This matches the time results of the corresponding Tasks 1 and 2. The hint for Task 1 was perceived less helpful for problem solving than the other hints (\cref{tab:hint_evaluation}) which might lead participants to stop trying early. It seems as if participants of \groupB remembered this and submitted incorrect programs early, again. On the contrary, however, the hint for Task 2 (\cref{fig:hint_example}) dealing with the bug pattern \emph{Missing Loop Sensing} was rated best (\cref{fig:hint_evaluation_quantitative}) and got more positive comments than the other hints (\cref{tab:hint_evaluation}). It may be that participants of \groupB were able to implement a solution quite fast, because they recognized the bug pattern \emph{Missing Loop Sensing} and could transfer a fix. 

On average, the effects are very similar for experienced and inexperienced participants (experienced: $\hat{A}_{12}$ = \processingTimeHigherExperienceEffectSizeMean; inexperienced: $\hat{A}_{12}$ = \processingTimeHigherNoExperienceEffectSizeMean). For experienced participants of \groupB, the time saving effects are again only larger for the Tasks 10 and 11.
%
%As for RQ1, being (in-)experienced seems to not hardly influence the time needed both for \groupB (experienced: \meanProcessingTimeExperienceTrmtHigher~minutes; inexperienced: \meanProcessingTimeNoExperienceTrmtHigher~minutes) and \groupA (experienced: \meanProcessingTimeExperienceCtrlHigher~minutes; inexperienced: \meanProcessingTimeNoExperienceCtrlHigher~minutes).
%
%
%However, there are some differences between the tasks. E.g. Tasks 10 and 11 are the only tasks of Tasks 8 to 14 where experienced participants benefit more---both in terms of functionality and time (\cref{tab:p-values_time_8-14}). 
%This indicates that the time needed depends only indirectly on the prior programming experience but is rather related to the functionality.

\summary{RQ2}{When debugging without hints, there are no significant differences between \groupB and \groupA. %participants who had received hints were not significantly more or less effective and efficient. %However, we observe a slight improvement.
%Participants who received hints in the past were generally not significantly more or less effective and efficient when having to debug without hints. However, we observe a slight improvement.
}
\newline

	%Our experiments suggest that hints have a small but positive effect on learning: For most of the tasks, the programs are repaired slightly more often and faster. Moreover, inexperienced participants benefit more from transferring hints---especially when programs are not too complex.}
	

%% file: content/discussion.tex
\section{Discussion}
\label{sec:discussion}

\begin{figure*}[tb]
	\centering
	\includegraphics[width=0.8\textwidth]{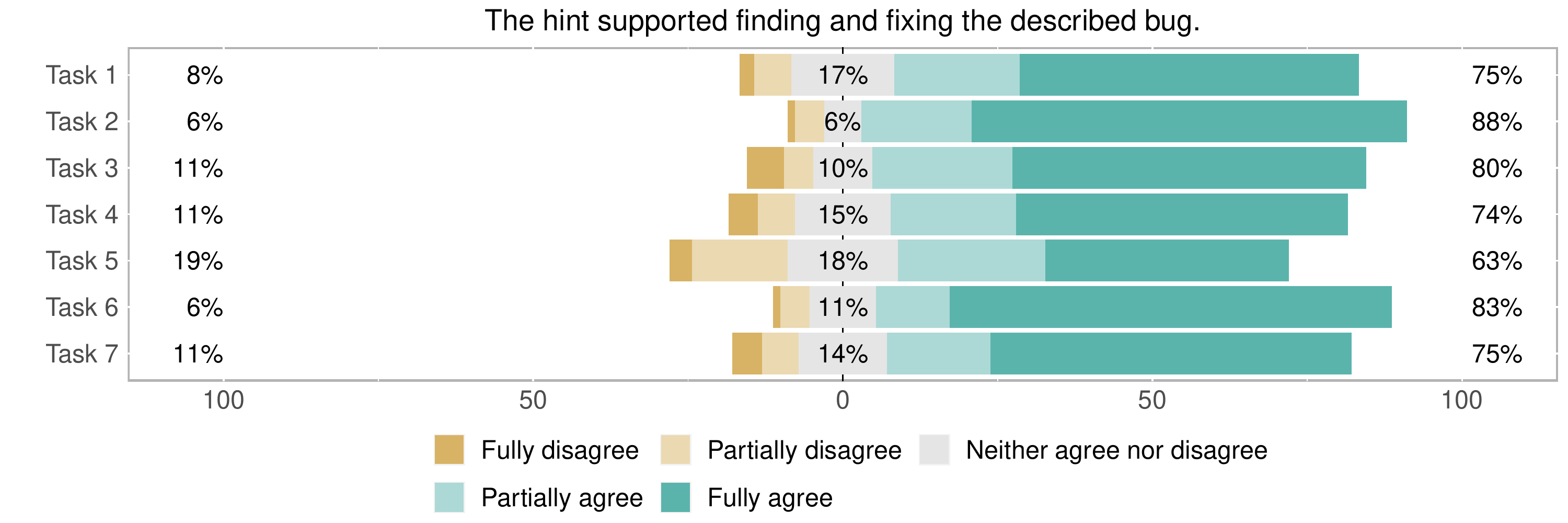} 
	\vspace{-0.5em}
	\caption{\label{fig:hint_evaluation_quantitative}Rating of the hints.}
%	\vspace{-0.5em}
\end{figure*}

\subsection{General Insights}
\subsubsection{RQ 1: Showing Hints}
The results on RQ 1 show that generic hints on bug patterns on average lead to debugging programs more often and faster.
%Effectiveness (RQ 1)
The increased effectiveness is important as teachers can only help the student to debug its program when understanding how to deal with the bug themselves. % This confirms that automatic hints can indeed support teachers with repairing student programs.
%Efficiency (RQ 1)
The increased efficiency implies that primary school teachers can react earlier to their student's problem---either by telling that they are not able to fix the bug or by trying to help the student to fix the bug. This is important as one teacher has to deal with many different problem-solving approaches at the same time~\cite{yadav2016expanding}. When hints are received it might be easier to deal with this challenge as less time is needed per bug. Still, when encountering hints for the first time, participants have to read, understand, and implement these hints, which also takes time. It needs further investigation to find out if seeing the same hints repeatedly in different programs, such that re-reading and comprehending does not need to be repeated, further reduces the amount of time needed.

We observed that more experienced participants benefit even slightly more from showing hints than inexperienced participants. Only for the Tasks 3 and 5 experienced participants benefit less in terms of effectiveness. The complexity of the program for Task 3 is very low (\cref{tab:task_complexity}) and experienced participants might not need the hint to locate the bug% and they might understand the problem of the bug pattern Comparing Literals even without the explanation of the problem (= first part of the hint text). However, the generic suggestion of the hint on how to remove the bug pattern (= second part of the hint text), has sometimes led to comprehension problems. It could therefore not always be implemented and also experienced participants stated: “I understood the problem but did not find the solution.” (P 66), “I solved it differently because I couldn’t find the required one right away.” (P 60) and “It was difficult for me to implement the solution despite the help!” (P 45)
. For Task 5, the task complexity does not stand out. However, its hint might harm the performance of experienced participants as they would have actually known where to look for the bug but were misled by the hint. However, for the five other tasks, experienced participants benefit more. This might be because basic knowledge is needed to understand the hints. Therefore, tool-support is especially useful for trained teachers in practice to face the challenge of multiple bugs at the same time.

% \textbf{Summarised, hints help to face the challenge of multiple bugs at the same time which is of special interest to teachers in practice.}

\subsubsection{RQ 2: Knowing Patterns}
%Effectiveness (RQ 2)
The results of RQ 2 do not indicate that learning opportunities with or without hints can be transferred significantly better. It seems that having repaired the corresponding bug pattern once does not suffice to transfer the hint's explanation. Still, it is an open question if repairing the same bug patterns more times would change the findings regarding the performance with vs. without hints. 
Another consideration is that the explanation is only displayed in the hint. Users could elaborate more on the hint if they additionally had to, e.g., describe it in their own words as suggested by Marwan et al.~\cite{marwan2019evaluation}. This might enhance the effects of hints on known patterns.

Our results suggest that inexperienced participants benefit slightly more from knowing patterns by the hints---apart from Task 10 and 11. This might be related to the relatively high complexity of these two programs (\cref{tab:task_complexity}), possibly overstraining inexperienced participants of both \groupA and \groupB. This suggests that knowledge of bug patterns can only support repairing programs correctly and save time when the programming experience suffices to understand a program of a given complexity level. Indeed, for all other tasks inexperienced participants benefit more. This might be because they need more direct instruction (e.g., provided by hints) to transfer their gained experiences than experienced participants. Consequently, bug patterns might be a useful concept to include in the curriculum for teachers in training.

\input{hint_evaluation_table.tex}

\subsection{Insights on Hint Generation}

\subsubsection{Location of the Bug%Tasks 5 and 12 (Message Never Received)
}
For both Tasks~5 and 12 the hint impairs effective and efficient debugging. They deal with the bug pattern \emph{Message Never Received} which has a peculiarity: While the hint is attached to the block sending a message, the fix requires modification of a different script which is often located in a different sprite, where that message should be handled. This is because static analysis only detects the pattern (e.g., that a message is sent but nowhere received) but it does not analyse content information (and thus does not return where the message should be received). This seems to provide too little help which matches the relatively bad rating of this hint (\cref{fig:hint_evaluation_quantitative}): The participants perceive the hint as  misleading and cognitively activating (\cref{tab:hint_evaluation}) which can also take time: \enquote{You had to think a little bit about it to finally get the correct result.} (P72). Furthermore, the hint appears to affect the performance for the similar bug in Task 12 negatively. Thus, hints generally need to make it very clear where the bug is located and where it can be repaired. Consequently, when changes have to be made in another script or even sprite than the shown one, this has to be highlighted especially in the hint to avoid misunderstandings.

%Effectiveness (RQ 1)

%\enquote{The bug has to be fixed in the sprite ``Friend'' and not in the sprite ``Coins'', therefore a somewhat misleading hint.} (P10). This results from hints on bug patterns always showing the location where the origin of the bug is located and not necessarily the bug itself. For Task 5, the origin is not synonymous to the bug itself as the sent message could be received in any sprite but only this sprite can not be detected automatically 
%\textbf{Consequently, when another sprite than the shown one has to be changed, this has to be highlighted in the hint to avoid misunderstandings.} %Thus, it is indispensable to evaluate and revise automatic feedback tools to ensure that e.g. hints are understandable.
%

\subsubsection{Provision of Several Solution Approaches%Task 1 and 8 (Message Never Sent)
}
%Effectiveness (RQ 1)
Showing hints also led to no significant improvement in effectiveness for Task 1 that deals with the bug pattern \emph{Message Never Sent}. Even though relatively many participants stated that the hint helped them to draw attention to important areas and to locate the bug, relatively few participants highlighted the support with problem solving (\cref{tab:hint_evaluation}). This might be explained by an obvious difference to other hints that misled some participants: The hint explains two possible solutions: (1) A message that is already sent in another script has to be selected or (2) a new message has to be sent in another script. Again, the automatic static analysis tools cannot detect whether an appropriate message is available (solution 1) or not (solution 2). The only other hint that explains two possible solutions is the hint on the bug pattern \emph{Missing Clone Initialisation}. However, the two provided solutions only differ in the block that is inserted and may therefore not be crucial. Consequently, attention must be paid if the results of the static analysis lead to several quite different solutions: Hints should then either provide information about which solution should be used in which context or at least mention that all solutions should be tried out until the intended output is reached.

%However, it is somewhat confusing that you should look for a suitable message in another script, as this can also be found in the shown script.}. 
%\textbf{This implies that when a bug pattern allows more than one solution, the hint has to describe in detail which script belongs to which solution.}

\subsubsection{Degree of Detail%Tasks 7 and 14 (Stuttering Movement)
}
%Effectiveness (RQ 1)
We observed the highest effects regarding the effectiveness of hints not only in Task~7 but also in the corresponding transfer Task 14. This is interesting, as the used bug pattern \emph{Stuttering Movement} certainly is more open to automatic refactoring than all other bug patterns of this study. Consequently the \litterbox hint presents itself quite customised and describes very detailed what should be done to fix the bug (\cref{tab:hints}). The classification of hint comments confirms that the hint was perceived as clear, important and least unnecessary (\cref{tab:hint_evaluation}). Maybe other bug patterns might not be so easily spotted, grasped and fixed by learners per se, due to greater variety of bug pattern instances or/and greater variety of appropriate fixing pathways and fixing outcomes. They consequently might need far more attention during learning than just trying to solve an individual task.
Compared to the effectiveness, however, the hint on the bug pattern \emph{Stuttering Movement} is slightly less helpful in terms of efficiency. This might be because %for the bug pattern \emph{Stuttering Movement} 
five blocks have to be changed---instead of one for the other bug patterns. Thus, reading or remembering the relatively long hint might take a while%---especially for less experienced participants
. 
In conclusion, hints need to provide clear information about how to remove the bug pattern.

%% file: hint_evaluation_table.tex
\begin{table}
\centering
\caption{Classification of the comments on the hints.}
\label{tab:hint_evaluation}
\vspace{-0.8em}
\setlength{\tabcolsep}{0.4em}
\begin{tabular}{llrrrrrrr}
\toprule
{} &  &T1 &  T2 &  T3 &  T4 &  T5 &  T6 &  T7 \\
\midrule
\multirow{9}{*}{\rotatebox[origin=c]{90}{positive}}    &            Problemsolving &      17 &      34 &      38 &      29 &      21 &      33 &      26 \\
 &                Clear hint &       8 &      14 &      13 &      16 &      16 &      19 &      20 \\
         &              Localisation &      20 &      14 &      12 &       9 &      10 &      11 &       5 \\
         &         Single components &      17 &      25 &       8 &      14 &       5 &       7 &       2 \\
         &         Attention drawing &      12 &      13 &       7 &       7 &       8 &       9 &       7 \\
        &        General assistance &       7 &       6 &       5 &       7 &       9 &       4 &       6 \\
         &                Importance &       6 &       7 &       7 &       6 &       0 &       1 &      10 \\
         &  Formulating feedback &       0 &       2 &       1 &       2 &       1 &       2 &       1 \\
         &                     TOTAL &      87 &     115 &      91 &      90 &      70 &      86 &      77 \\
\midrule
\multirow{3}{*}{\rotatebox[origin=c]{90}{neutral}} &      Cognitive activation &       3 &       4 &       2 &       5 &      11 &       3 &       5 \\
         &        Partial usefulness &       0 &       1 &       1 &       2 &       0 &       0 &       1 \\
         &                     TOTAL &       3 &       5 &       3 &       7 &      11 &       3 &       6 \\
\midrule
\multirow{7}{*}{\rotatebox[origin=c]{90}{negative}} &                Misleading &       3 &       1 &       0 &       6 &      24 &       1 &       3 \\
         &    Comprehension problems  &       3 &       2 &       7 &       6 &       8 &       2 &       7 \\
         &   Insufficient assistance &       5 &       1 &       3 &       3 &       3 &       1 &       6 \\
         &        No problem solving &       0 &       1 &       3 &       3 &       3 &       1 &       7 \\
         &                      Time &       2 &       0 &       5 &       2 &       1 &       0 &       1 \\
         &   No deeper understanding &       0 &       3 &       0 &       1 &       0 &       2 &       3 \\
         &                     TOTAL &      13 &       8 &      18 &      21 &      39 &       7 &      27 \\
\midrule
\multirow{4}{*}{\rotatebox[origin=c]{90}{unnecessary}} &     Generally unnecessary &       3 &       3 &       4 &       3 &       4 &       1 &       1 \\
         &       Independent solving &       3 &       1 &       1 &       0 &       4 &       2 &       1 \\
         &      Alternative solution &       0 &       0 &       3 &       0 &       0 &       1 &       0 \\
         &                     TOTAL &       6 &       4 &       8 &       3 &       8 &       4 &       2 \\
\bottomrule
\end{tabular}
\end{table}

%% file: content/conclusions.tex
\section{Conclusions}
\label{sec:conclusions}

Teachers are faced with bugs and have to be able to identify and repair them. In this paper, we empirically evaluated to what extent hints can support primary school teachers in training.
Our experiment confirmed that hints can help repairing bugs in students' programs, and we found no evidence that the hints would impede the learning effects gained from debugging.
While this result is encouraging for research on hint generation, the observations and the participants' evaluation of the hints provided valuable insights on how to further improve hints related to misconceptions and automated hint generation in future research. 

While we focused on teachers as target for the hints, in the future we also plan to investigate creating hints specifically for younger children. These hints would have to be kept very simple and yet point at an aspect of the underlying problem in an adequate manner. However, hints aimed at teachers have the potential to indirectly reach a greater variety of students, since the insight encapsulated in these hints can be transformed and tailored by the teachers to their students' needs.  
%
%In this paper we focused on effectiveness in terms of functionality and efficiency in terms of time. 
As a next step we plan to analyse the output of this transformation in a fictitious student setting. 
A further important aspect of automatically generated hints is that tools may produce false positives, and not all bugs in programs will be instances of bug patterns. We plan to investigate the effects of these issues on the debugging performance, and hope that insights will also help to improve hint generation techniques.
%

%The skills and knowledge gained from these hints can be transferred only limitedly to similar programs without hints. This indicates that hints should be given permanently and not only once.
%
%
%The hints showing directly where the bug is located was perceived particularly well by the participants. Therefore hints should always be attached to the respective block.
%However, the instant appearance of the hints was partially not considered useful. This can be avoided by only giving the hint when requested or after a certain amount of time.
%The participants' attitude towards the accuracy of the hints varied a lot. Some participants were in favour of the hint not revealing the whole solution and considered it precisely formulated, whereas others did not understand the hint and considered it too inaccurate. This leads to the idea of providing different versions of one hint. Effects of differentiated hints can then in a next step be examined.
%

%In this paper we focused on the bug patterns \textit{Position Equals Check} and \textit{Missing Loop Sensing}. Giving hints about these bug patterns was found to be advantageous when debugging student programs. Effects of hints about other common bug patterns should be further examined to see if hints are more suitable for some bug patterns than for others.